\def\epsfpreprint{Y}   
                       
                       \if \epsfpreprint Y
\documentstyle[12pt,epsf]{article} \fi \if \epsfpreprint N
\documentstyle[12pt]{article} \fi

\def\figsizeA{6.4}
\def\figsizeB{4.2}
\def\figsizeC{7.2}

\def\figure#1#2#3{\if \epsfpreprint Y \epsfxsize=#3truein
\centerline{\epsffile{fig_#1.eps}}
\centerline{\vbox{{\bf \noindent Figure #1.} #2}}
\bigskip \fi}

\def\Psibar{\overline{\Psi}}
\def\psibar{\overline{\psi}}

\def\slash{\!\!\!\!/}
\def\platslash{\overline p \!\!\!/} 
\def\plat{\overline p}
\def\meff{m_{\rm eff}}

\def\spose#1{\hbox to 0pt{#1\hss}}
\def\ltapprox{\mathrel{\spose{\lower 3pt\hbox{$\mathchar"218$}}
 \raise 2.0pt\hbox{$\mathchar"13C$}}}
\def\gtapprox{\mathrel{\spose{\lower 3pt\hbox{$\mathchar"218$}}
 \raise 2.0pt\hbox{$\mathchar"13E$}}}
\def\inapprox{\mathrel{\spose{\lower 3pt\hbox{$\mathchar"218$}}
 \raise 2.0pt\hbox{$\mathchar"232$}}}

\def\one{Observables (for definitions see section \ref{sec-observables}) 
along the configuration space trajectory, eq. \ref{U1_traj}, labeled
by $\tau$.  The parameters have values: $L=6$, $\mu l = 3.0$, $m_0 =
0.9$, $m_f=0$, $L_s = 14$.}
\def\two{Observables (for definitions see section \ref{sec-observables}) 
along the configuration space trajectory, eq. \ref{U1_traj}, labeled
by $\tau$.  The parameters have values: $L=6$, $\mu l = 3.0$, $m_0 =
0.9$, $m_f=0$, $L_s = 4,6,8,10,12,14,\infty$. At $L_s=\infty$ $P_1=0$.}
\def\three{$P_1$ from fig. 2 vs. $L_s$ at different values of
$\tau = 0$, $0.3$, $0.4$, $0.55$ corresponding to diamonds, squares,
crosses, octagons.}
\def\four{Expectation value of the chiral condensate in a 
configuration space restricted on the trajectory of eq. \ref{U1_traj}
vs. $L_s$ for different gauge couplings. The parameters have values:
$L=6$, $m_0 = 0.9$, $m_f=0$, $\mu l = 3.0, 2.5, 2.0, 1.5$, and $1.0$
corresponding to diamonds, squares, crosses, octagons, and stars.}
\def\five{$P_1$ along the configuration space trajectory, 
eq. \ref{U1_traj}, labeled by $\tau$.  The parameters have values:
$L=6$, $\mu l = 3.0$, $m_0 = 0.9$, $m_f=0.1$, $L_s =
4,6,8,10,12,14,16,\infty$.}
\def\six{$\left[ <\psibar \psi>_{L_s} - 
<\psibar \psi>_{\infty} \right] / <\psibar \psi>_{L_s}$ vs. $L_s$ for
$m_f=0.1$. $<\psibar \psi>$ is the chiral condensate in a
configuration space restricted on the trajectory of eq. \ref{U1_traj}.
The parameters have values: $L=6$, $\mu l = 3.0$, and $m_0=0.9$.}
\def\seven{$<\psibar \psi>$ vs. $L_s$ calculated
in an ensemble of configurations generated
by applying small fluctuations of size $\epsilon$ to the trivial
configuration ($U = e^{i r\pi}$ with $r$ randomly distributed in
$-\epsilon < r <\epsilon$). The parameters have values: $L=6$, $m_0 =
0.9$, $m_f=0$, $\epsilon = 0.4, 0.3, 0.2, 0.1$ and $0.01$
corresponding to octagons, stars, squares, crosses, and diamonds.}
\def\eight{$<w>/m_\gamma^2$ vs. $m_f$ for $L=6$, $m_0 =0.9$, $\mu l = 3.0$ 
using the Overlap. The dotted lines are the $m_f=0$ result $\pm$ the
corresponding statistical error.}
\def\nine{$<\psibar \psi>/ m_\gamma$ vs. $m_f$ for
$L=6$, $m_0 =0.9$, $\mu l = 3.0$ using the Overlap. The fits are to
$<\psibar \psi> = A m_f^p$.  Both fits have a $\chi^2$ per degree of
freedom of about one.  For $m_f < 0.1$ $p = 0.996(3)$, while for $m_f>0.1$ 
$p = 0.32 (2)$.}
\def\ten{Time history of $w$, eq. \ref{pbp2}, for six different values of 
$m_f$. Notice the different scale of the $m_f=0.05$ and $m_f=0.0$ graphs.
The parameters are $L=6$, $\mu l = 3.0$, $m_0=0.9$ and $L_s=14$.}
\def\eleven{$<\psibar \psi> / m_\gamma$, eq. \ref{pbp1}, vs. $L_s$ for $m_f=0$, 
fixed physical volume $\mu l = 3.0$, $m_0 = 0.9$ and for four
different lattice spacings $\mu l / L = \mu a$, with $L= 6, 8, 10, 12$
corresponding to the lines from top to bottom.}

\def\twoelve{The exponentiated slopes, $e^{-c}$,  
of the lines in figure $11$ vs. $1/L \sim a$. 
The diamonds correspond to the faster rates $6 \leq L_s \leq 10$ while
the crosses to the slower ones $12 \leq L_s \leq 22$.}
\def\thirteen{$[<\psibar \psi> / m_\gamma]^3$ vs. $L_s$ for four different $m_f$.
The physical volume is fixed at $\mu l = 3.0$, $m_0=0.9$ and the
lattice spacing is set by $L=6$.  The fits are to a function $A + B
e^{-c L_s}$. The dotted line is the $L_s=\infty$ result $\pm$ the
error from figure $9$. The cross is the coefficient $A$.}
\def\fourteen{$<w> / m_\gamma^2$ vs. $L_s$ for four different $m_f$.
The physical volume is fixed at $\mu l = 3.0$, $m_0=0.9$ and the
lattice spacing is set by $L=6$.  The fits are to a function $A + B
e^{-c L_s}$. The dotted line is the $L_s=\infty$ result $\pm$ the
error from figure $8$. The cross is the coefficient $A$.}
\def\fifteen{$[<\psibar \psi> / m_\gamma]^3$ vs. $L_s$ for four different 
lattice spacings set by $L$ at fixed $m_f=0.2$.  The physical volume
is fixed at $\mu l = 3.0$ and $m_0=0.9$. The fits are to a function $A
+ B e^{-c L_s}$. The dotted line is the $L_s=\infty$ result $\pm$
the error. The cross is the coefficient $A$.}
\def\sixteen{$<w> / m_\gamma^2$ vs. $L_s$ for four different 
lattice spacings set by $L$ at fixed $m_f=0.2$.  The physical volume
is fixed at $\mu l = 3.0$ and $m_0=0.9$. The fits are to a function $A
+ B e^{-c L_s}$. The dotted line is the $L_s=\infty$ result $\pm$
the error. The cross is the coefficient $A$.}
\def\seventeen{$[<\psibar \psi> / m_\gamma]^3$ vs. $L_s$ for two different 
lattice spacings set by $L=4$ (squares) and $L=10$ (diamonds) at
$m_0=0.9$.  The physical volume and $m_f L$ are fixed at $\mu l =
3.0$ and $m_f L = 2.0$. The fits are to  $A + B e^{-c
L_s}$. The dotted lines are the $L_s=\infty$ results $\pm$ the error.
The cross is the coefficient $A$.}

\def\eighteen{$<w> / m_\gamma^2$ vs. $L_s$. The parameters are as in
figure $17$.}
\begin{document}

\title{\bf Chiral Symmetry Restoration in the Schwinger Model with 
Domain Wall Fermions}
\vskip 1. truein
\author{Pavlos M. Vranas \\
Columbia University \\ 
Physics Department \\ 
New York, NY 10027 }

\maketitle

\vskip -4.0 truein
\rightline{CU-TP-838}
\rightline{To appear in Phys. Rev. D.}
\vskip 5.3 truein

\begin{abstract}

Domain Wall Fermions utilize an extra space time dimension to provide
a method for restoring the regularization induced chiral symmetry
breaking in lattice vector gauge theories even at finite lattice
spacing. The breaking is restored at an exponential rate as the size
of the extra dimension increases. Before this method can be used in
dynamical simulations of lattice QCD, the dependence of the
restoration rate to the other parameters of the theory and, in
particular, the lattice spacing must be investigated. In this paper
such an investigation is carried out in the context of the two flavor
lattice Schwinger model.

\end{abstract}

\newpage

\section{Introduction}
\label{sec-introduction}

When fermions are discretized on a d-dimensional lattice they
``double'' producing $2^d$ species for each flavor. In order to remove
the unwanted degrees of freedom special care must be taken.  For a
vector theory, like QCD, two methods have been used to deal with this
problem, but both break the global symmetries of the continuum theory.
Wilson fermions \cite{wilson-fermions} are implemented by adding
an irrelevant operator to the action.  This operator makes all but
one of the species heavy (with masses close to the cutoff).  For the
$N_f$ flavor QCD this operator breaks the $SU(N_f)_L \times SU(N_f)_R$
chiral symmetry down to $SU(N_f)$. This explicit breaking is severe
and requires fine tuning of the bare quark mass in order to obtain a
massless theory. Even then the size of the breaking is proportional to
the lattice spacing and only close to the continuum limit the explicit
breaking becomes small.  Staggered fermions
\cite{staggered-fermions} break the $SU(N_f)_L \times SU(N_f)_R$
chiral symmetry down to $U(1) \times U(1)$. Because of the remnant
of chiral symmetry the massless theory can be reached by simply taking
the bare quark mass to zero.  However, the flavor symmetry of the
theory has been compromised and is also only recovered as the
continuum limit is approached.

Despite these problems both methods have been very successful in
describing the light hadron spectrum at zero temperature. However,
both methods have difficulties in studying the finite temperature
phase transition. Wilson fermions have a complicated phase diagram
that, at the presently accessible lattice spacings, makes it hard to
extract the relevant physics. Staggered fermions, because of the exact
remnant of chiral symmetry, do not suffer from this problem. However,
at the presently accessible lattice spacings, the breaking of flavor
symmetry makes two of the three pions heavy.  This can have important
physical consequences since the transition temperature is of the order
of the pion mass. For a review on the finite temperature phase
transition with both types of fermions the reader is referred to
\cite{Ukawa-review} and references therein.

A few years ago a new method for discretizing fermions was developed in
order to address the more difficult problems associated with chiral
gauge theories \cite{Kaplan}.  In the following years this method was
further developed (see \cite{DWF-reviews} and references therein) with
important progress in the development of chiral gauge theories
\cite{NN1}, \cite{NN2}.  The basic idea follows from the fact that a massive
vector theory in $2n+1$ dimensions, with a mass term that changes sign
along the $2n+1$ dimension, develops a massless chiral zero mode that
is exponentially bound along the $2n+1$ direction to the region where
the mass changes sign.  From the point of view of the $2n$ dimensional
world this is a chiral fermion. This region is called Domain Wall and
this type of fermion is called Domain Wall Fermion (DWF). When such a
theory is discretized species doubling also occurs. However, since the
$2n+1$ dimensional theory is vector-like the extra species can be
removed with the addition of a standard Wilson term. The resulting
theory has a single chiral fermion exponentially bound to the wall.
If, for practical reasons, the $2n+1$ dimension is made finite with
periodic boundary conditions for the mass then the mass must change
sign one more time. In that region (anti-wall) an exponentially bound
chiral zero mode with opposite chirality appears. As a result, the
theory becomes vector-like.  Different types of boundary conditions
yield similar problems. In order to preserve the single chiral mode,
the $2n+1$ dimension must be kept infinite. At first sight this may
seem impractical. However, Narayanan and Neuberger developed a method,
called the Overlap formalism, that makes it possible to deal with this
infinity \cite{NN1}.

The Overlap formalism develops a transfer matrix along the $2n+1$
dimension and an associated Hamiltonian.  The gauge fields are defined
only on the $2n$ dimensional space and are taken to be independent of
the $2n+1$ coordinate \cite{NN1}. In essence, the extra dimension is
treated as a complicated flavor space. The resulting formalism
involves two Hamiltonians, one for the region of positive mass and one
for the region of negative mass.  The chiral determinant is the
determinant of the overlap of the two ground states associated with
each Hamiltonian and it can be calculated explicitly once all the
negative eigenvectors of both Hamiltonians are known. For a finite
$2n$ dimensional lattice the Hamiltonians are finite size matrices of
size $\sim V \times V$ where $V$ is the $2n$ dimensional volume and
their eigenvectors can be readily calculated.  The resulting chiral
determinant has the correct magnitude and a phase that exhibits the
correct gauge dependence for ``smooth'' gauge fields.  For ``rough''
gauge fileds the phase exhibits a mild breaking of gauge symmetry even
for anomaly free theories. This problem has been resolved in
\cite{NN2}.

These methods can also be used to formulate a vector theory.  In this
case the boundary conditions along the $2n+1$ dimension are set to be
periodic. A Dirac fermion emerges with the positive chirality
component bound on the wall and the negative chirality bound on the
anti-wall. If the $2n+1$ dimension is taken to be infinite the two
chiralities are decoupled and the resulting theory has intact chiral
symmetries! Again, this infinity can be dealt with the Overlap
formalism and now there are no issues associated with the phase of the
determinant since, for a vector theory, the determinant is real.
Therefore, the Overlap formalism provides an ideal lattice
regularization of vector theories where the chiral symmetries are left
intact even for finite lattice spacing. Also, the anomalous breaking
of the axial symmetry is reproduced in an elegant way along with a
formula for the index of the chiral Dirac operator
\cite{NN1}.

The Overlap formalism was used in a dynamical numerical simulation of the
massless single and multiflavor Schwinger model with good results
\cite{NNV}. Numerical simulations of QCD using the overlap formalism
would clearly be very appealing. However, as mentioned above, such a
simulation would require the calculation of all negative eigenvectors
of matrices of size $\sim V \times V$. This makes such a
calculation prohibitive for present generation supercomputers.

An obvious alternative (see \cite{DWF-reviews} and references therein)
is to keep the $2n+1$ dimension finite and use standard Hybrid Monte
Carlo type algorithms to simulate the theory in $2n+1$ dimensions.  Of
course, the exact chiral symmetry will be spoiled but it will be
recovered as the size $L_s$ of the $2n+1$ dimension is sent to
infinity. Therefore, even at finite lattice spacing one can control
the restoration of the regularization induced chiral symmetry breaking
by using the parameter $L_s$. This involves no fine tuning and,
furthermore, since the two chiralities decay exponentially away from
the wall (anti-wall), one would expect that the restoration of chiral
symmetry would be exponential i.e.  $\sim e^{-c L_s}$, $0 < c$.  The
computer cost of such a simulation would be $L_s$ times larger than a
simulation of standard Wilson fermions with the same physical masses.
Since for present day supercomputers a value of $L_s$ greater than $10
- 20$ will make simulations impractical, an important question to ask
is what is the rate ``$c$'' of restoration of chiral symmetry and how
does it depend on the other parameters of the theory and in particular
on the lattice spacing. In \cite{Shamir},
\cite{Furman-Shamir} some of the issues relating to this question
were investigated analytically. Numerical work in \cite{Jaster},
\cite{PMV}, \cite{Blum-Soni} has yielded encouraging results 
and in particular the interesting work of \cite{Blum-Soni}
indicates that DWF can successfully address problems related
to the evaluation of weak matrix elements. However, these works have
only marginally addressed this particular question.  
Before full scale dynamical
QCD simulations are performed this question should be answered. In
this paper this question is investigated in the context
of the two flavor lattice Schwinger model.

A useful variation of the wall, anti-wall model studied in
\cite{BDF} was proposed in the context of vector lattice gauge
theory in \cite{Shamir}, \cite{Furman-Shamir}. There, instead of
having a mass that changes sign in two places along the $2n+1$
dimension (say at $0$ and $L_s/2$), the mass is kept fixed to some
positive value $m_0$, but the boundary conditions are taken to be free
at the ends of the $2n+1$ dimension.  Again, two zero modes with
opposite chiralities emerge, but they are now bound at the opposite
ends of the $2n+1$ dimension and are separated by a distance $L_s$
rather than $L_s/2$ as in the original model.  Therefore, the
expectation is that for the same $L_s$ this model will achieve better
restoration of chiral symmetry.

Another feature added in this model is the introduction of an explicit
chiral symmetry breaking term that connects the two ends with strength
$m_f$. This gives mass to the fermion in addition to the one
resulting because of the finite extend $L_s$.  The reason for adding
this term is that it provides linear control over the fermion mass
instead of the exponential one provided by $L_s$. Furthermore, in a
numerical computation, it makes much more sense to vary $m_f$ rather
than $L_s$ in order to control the mass. Therefore, for a given $m_f$
one would like to keep $L_s$ large enough so that it does not affect
the fermion mass in any significant way. This method and the
associated Overlap formalism will be used throughout this paper.

The theory has five parameters.  The first two are the lattice spacing
$a$ and the physical extent $l$ along one direction of the $2n$
dimensional box (they are controlled by the pure gauge coupling $g_0$
and the size in lattice units $L$).  The remaining three parameters
$m_0$, $L_s$ and $m_f$ all control, to some extent, the amount of chiral 
symmetry breaking and therefore the effective fermion mass. For $L_s
\rightarrow \infty$ the theory is chirally symmetric except for the 
explicit breaking introduced by $m_f$.  As a result, the effective
fermion mass vanishes linearly with vanishing $m_f$
\cite{Furman-Shamir}. But for finite $L_s$ this is not the case. As
mentioned above, even for $m_f=0$ the restoration of chiral symmetry
is expected to be exponential $\sim e^{-c L_s}$.

One would expect that the exact continuum solution of the Schwinger
model would be useful to compare with results obtained on the
lattice.  Unfortunately, this is only partially true. The
regularization with DWF introduces to the two dimensional action a
four-Fermi term with some coefficient. Since, for the two dimensional
model, this operator is not irrelevant, the continuum theory will be
different from a continuum theory with no four-Fermi term.  Although
the continuum theory has been solved with such a term present
\cite{Dettki-Sachs-Wipf}, the results can not be directly compared
since the value of the coefficient arising from the DWF has not been
calculated.  This problem was encountered in \cite{NNV} and \cite{NN2}
and made the comparison with continuum results complicated.
Fortunately, for the purposes of this work, the continuum results are
not needed. In fact, there is a much more relevant comparison that can
be made.  At every step the value of any observable at finite $L_s$
can be directly compared with its value at infinite $L_s$ calculated
using the Overlap on the same lattice size and lattice spacing.

The paper is organized as follows: 
In section \ref{sec-model} the model and the
corresponding Overlap implementation is reviewed.
In section \ref{sec-observables} the definitions of the various
observables used in this paper are given.
In section \ref{sec-free} the free theory for finite $L_s$ is
discussed, the full expression for the propagator is given and the
``effective'' bare fermion mass is identified.  
In section \ref{sec-interact} some general considerations
regarding the interacting theory are presented. These
considerations lead to specific predictions.
In section \ref{sec-overlap-sim} the results of a full dynamical
simulation using the Overlap with non zero mass are presented. These
results, interesting in their own right, are used to compare with the
finite $L_s$ results of the next section.
Section \ref{sec-hmc} describes the results of a dynamical simulation
of the $2+1$ dimensional system for various values of the parameters
on a fixed physical volume. The algorithm used is a standard Hybrid
Monte Carlo (HMC) algorithm.  The numerical results confirm the
predictions made in section
\ref{sec-interact} and together outline the mechanisms of chiral
symmetry restoration in the model.
Section \ref{sec-conclusions} contains a summary and conclusions.


\section{The Model}
\label{sec-model}

In this section the model and the corresponding Overlap
formalism \cite{NN1} implementation 
is reviewed for the benefit of the reader and in
order to establish notation \cite{Shamir},
\cite{Furman-Shamir}. The following is for a single flavor. The
generalization to more flavors is straightforward.

The partition function of the single flavor $2n+1$ dimensional model
is:
\begin{equation}
Z = \int [dU] \int [d\Psibar d\Psi] \int [d\Phi^\dagger d\Phi] e^{-S}
\label{Z}
\end{equation}
$U_\mu(x)$ is the gauge field, $\Psi(x,s)$ is the fermion field and
$\Phi(x,s)$ is a bosonic Pauli Villars (PV) type field. $x$ is
a coordinate in the $2n$ dimensional space-time box with extent $L$
along each of the directions, $\mu = 1,2,\dots 2n$, and $s=0,1, \dots,
L_s-1$, where $L_s$ is the size of
the $2n+1$ direction and is taken to be an even number.  The action
$S$ is given by:
\begin{equation}
S = S(g_0, L, L_s, m_0, m_f) = S_G(U) + S_F(\Psibar, \Psi, U) +
S_{PV}(\Phi^\dagger, \Phi, U)
\label{action}
\end{equation}
where:
\begin{equation}
S_G = {1\over g_0^2} \sum_p Re Tr[I - U_p]
\label{action_G}
\end{equation}
is the standard plaquette action with $g_0$ the lattice gauge
coupling. In this paper the coupling $g_0$ is exchanged for the
parameter:
\begin{equation}
(\mu l) = {g_0 \over \sqrt{\pi} } L
\label{mu_l}
\end{equation}
where $l$ is the physical size of the $2n$ dimensional box along one of
its dimensions and $\mu$ is a mass related to the photon mass with:
\begin{equation}
m_\gamma = \sqrt{N_f} \mu
\label{m_gamma}
\end{equation}
where $N_f$ is the number of flavors. With these choices $\mu l$ is
the physical box size in units of $\mu$ and $\mu l / L= \mu a$ is
the lattice spacing in units of $\mu$.

The fermion action is:
\begin{equation}
S_F = - \sum_{x,x^\prime,s,s^\prime} \Psibar(x,s) D_F(x,s; x^\prime,
s^\prime) \Psi(x^\prime,s^\prime)
\label{action_F}
\end{equation}
with the fermion matrix given by:
\begin{equation}
D_F(x,s; x^\prime, s^\prime) = \delta(s-s^\prime) D\slash(x,x^\prime)
+ D\slash^\bot(s,s^\prime) \delta(x-x^\prime)
\label{D_F}
\end{equation}
\begin{eqnarray}
D\slash(x,x^\prime) &=& {1\over 2} \sum_\mu\left[ (1+\gamma_\mu)
U_\mu(x) \delta(x+\hat\mu - x^\prime) + (1-\gamma_\mu)
U^\dagger_\mu(x^\prime) \delta(x^\prime+\hat\mu - x) \right] \nonumber \\
&+& (m_0 - 2n)\delta(x-x^\prime)
\label{Dslash_F}
\end{eqnarray}
\begin{equation}
D\slash^\bot(s,s^\prime) = \left\{ \begin{array}{ll} P_R
\delta(1-s^\prime) - m_f P_L \delta(L_s-1 - s^\prime) - \delta(0-
s^\prime) & s=0 \\ P_R \delta(s+1 - s^\prime) + P_L \delta(s-1 -
s^\prime) - \delta(s-s^\prime) & 0 < s < L_s-1 \\ -m_f P_R
\delta(0-s^\prime) + P_L \delta(L_s-2 - s^\prime) - \delta(L_s-1 -
s^\prime) & s = L_s -1
\end{array}
\right. 
\label{Dslash_perp_f}
\end{equation}
\begin{equation}
P_{R,L} = { 1 \pm \gamma_5 \over 2}
\end{equation}
where $m_0$ is a $2n+1$ dimensional mass representing the ``height'' of
the Domain Wall. In order for the doubler species to be removed one
must set $0<m_0<2$ \cite{Kaplan}. However, this range is further
restricted by the requirement that the transfer matrix along the $2n+1$
direction be positive \cite{NN1}:
\begin{equation}
0 < m_0 < 1
\label{m0_range}
\end{equation}
The gamma matrices are taken in the chiral basis
and are the same as in the last reference in \cite{NN1}.  In two
dimensions they are:
\begin{equation}
\gamma_1 = \left( \begin{array}{cc} 0 & 1 \\ 1 & 0 \end{array} \right), \ \  
\gamma_2 = \left( \begin{array}{cc} 0 & i \\ -i & 0 \end{array} \right), \ \ 
\gamma_5 = \left( \begin{array}{cc} 1 & 0 \\ 0 & -1 \end{array} \right) 
\end{equation}
The PV action is designed to cancel the contribution of the heavy
fermions in the large $L_s$ limit.  This is necessary because the
number of heavy fermions is $\sim L_s$ and at the $L_s \rightarrow
\infty$ limit they produce bulk type infinities \cite{NN1}. There is
some flexibility in the definition of the PV action since different
actions could have the same $L_s \rightarrow \infty$ limit.  However,
the choice of the PV action may affect the approach to the $L_s
\rightarrow \infty$ limit.  A slightly different action than the one
used in \cite{Furman-Shamir} is used here. This action is easier to
implement numerically and for finite $L_s$ it projects the ground
state of the transfer matrix $T$ better; the projector is $T^{L_s}$
instead of $T^{L_s/2}$ (see below). Also, even for finite $L_s$, 
it exactly cancels out the
fermion action when $m_f = 1$ resulting into a pure gauge theory.  The
PV action is:
\begin{equation}
S_{PV} =
\sum_{x,x^\prime,s,s^\prime} \Phi^\dagger(x,s) 
D_F[m_f=1](x,s; x^\prime, s^\prime) \Phi(x^\prime,s^\prime)
\label{action_PV}
\end{equation}

The transfer matrix along the $2n+1$ direction for this model is:
\begin{equation}
T = e^{- \hat a^\dagger H \hat a}
\label{transf_mat}
\end{equation}
where $\hat a^\dagger, \hat a$ are creation and annihilation operators
that obey canonical anticommutation relations and span a Fock space
with vacuum state $|0>$. These operators live on the sites of the $2n$
dimensional lattice and carry spin, color and flavor indices. The
left/right component decomposition of $\hat a$ is:
\begin{equation}
\hat a = \left( \begin{array}{c} \hat c \\ \hat d^\dagger \end{array} \right)
\label{a_hat}
\end{equation}
The single particle Hamiltonian $H$ is defined by:
\begin{equation}
e^{-H} = \left( \begin{array}{cc} B^{-1} & B^{-1} C \\ C^\dagger
B^{-1} & C^\dagger B^{-1} C + B
\end{array}
\right)
\label{Hamiltonian}
\end{equation}
\begin{equation}
B(x,y) = {1\over2} \sum_{\mu=1}^{2n}  
\left[ 2 - U_\mu(x) \delta(x+\hat\mu - y) 
- U^\dagger_\mu(y) \delta(y + \hat\mu - x)
\right]
+ (1 - m_0) \delta(x-y)
\label{B}
\end{equation}
\begin{equation}
C(x,y) = {1\over2} \sum_\mu\left[U_\mu(x) \delta(x+\hat\mu - y) -
U^\dagger_\mu(y) \delta(y + \hat\mu - x) \right] \sigma_\mu
\label{C}
\end{equation}
with $\sigma_1 = 1, \sigma_2 = i$ in two dimensions.

The fermionic and PV effective actions can be expressed in terms of the
transfer matrix as:
\begin{equation}
e^{-S^F_{\rm eff}[L_s]} =
\det(D_F[m_f]) = \det(B)^{L_s} Tr\left[ T^{L_s} {\cal O}(m_f)\right]
\label{det_F}
\end{equation}
\begin{equation}
e^{-S^{PV}_{\rm eff}[L_s]} =
\left( \det(D_F[m_f=1]) \right)^{-1}
= \left( \det(B)^{L_s} {Tr\left[ T^{L_s}\right]} \right)^{-1}
\label{det_PV}
\end{equation}
where the operator ${\cal O}(m_f)$ implements the boundary conditions
and contains all the $m_f$ dependence:
\begin{equation}
{\cal O}(m_f) = \prod_n (\hat c_n \hat c_n^\dagger + m_f \hat
c_n^\dagger \hat c_n) (\hat d_n \hat d_n^\dagger + m_f \hat
d_n^\dagger \hat d_n)
\label{Omf}
\end{equation}
For $m_f=1$ it is the identity operator and for $m_f=0$ is a
projection operator to a state $|0^\prime>$:
\begin{equation}
|0> = \prod_n \hat d^\dagger |0^\prime>
\label{0_prime}
\end{equation}

In the infinite $L_s$ limit $T^{L_s}$ becomes a projection operator to
the ground state of $- \hat a^\dagger H \hat a$,
\begin{equation}
\lim_{L_s \rightarrow \infty} T^{L_s} \rightarrow e^{-\lambda_0} |0_H><0_H|
\label{Tprojector}
\end{equation}
where $|0_H>$ and $\lambda_0$ are the ground state eigenvector and
eigenvalue of $\hat a^\dagger H \hat a$ obtained by filling all
negative energy states.  To get an explicit relation between $|0_H>$
and $|0>$ let $R$ be the eigenvector matrix of the single particle
Hamiltonian $H$. The matrices $H$ and $R$ have size $N \times N$ where
$N = {\rm spin} \times {\rm color} \times {\rm flavor} \times V$.
$R$ can be put in the form:
\begin{equation}
R = \left( \begin{array}{cc} P^- & P^+ \\ Q^- & Q^+ \end{array}
\right)
\label{R}
\end{equation}
where the rows labeled $P$ correspond to the left chirality states
(with creation/annihilation operators $\hat c^\dagger, \hat c$) and the
rows labeled $Q$ correspond to the right chirality states (with
creation/annihilation operators $\hat d^\dagger, \hat d$). The $\pm$
splitting of the columns corresponds to eigenvectors with
positive/negative eigenvalues.  With $N^{\pm}$ denoting the number of
positive/negative eigenvalues the size of the $P^-, Q^-$ matrices is
$N/2 \times N^-$ and the size of the $P^+, Q^+$ matrices is $N/2
\times N^+$. Then it can be shown that:
\begin{equation}
|0_H> = \prod_{i=1}^{N^-}(\hat c_{l_i}^\dagger P_{{l_i},i}^- + \hat
d_{l_i} Q_{{l_i},i}^- )|0>
\label{0_H}
\end{equation}
From equations \ref{det_F}, \ref{det_PV} and \ref{Tprojector} the
effective action for the fermion and PV fields in the $L_s \rightarrow \infty$
limit is given by the Overlap formula:
\begin{equation}
e^{-S_{\rm eff}[L_s = \infty, m_f]} = e^{-S^F_{\rm eff}[L_s = \infty,
m_f] -S^{PV}_{\rm eff}[L_s = \infty]} = <0_H| {\cal O}(m_f) |0_H>
\label{Overlap_seff}
\end{equation}
For $m_f=1$, $<0_H| {\cal O}(1) |0_H> = 1$, corresponding to a
theory with no fermions.  The $m_f=0$ case corresponds to massless
fermions and the Overlap takes the special form:
\begin{equation}
e^{-S_{\rm eff}[L_s = \infty, m_f=0]} = \left| <0_H|0^\prime>\right|^2
\label{Overlap_mf0}
\end{equation}
It can be shown that:
\begin{equation}
\left| <0_H|0^\prime>\right|^2 = 
\left| \det( Q^-) \right|^2
\label{det_Q}
\end{equation}
If $N^- = N/2$, $Q^-$ is a square matrix and the Overlap will in
general be non zero.  However, if $N^- \neq N/2$ then $Q^-$ is not a
square matrix and its determinant is identically zero.  From equations
\ref{0_prime}, \ref{0_H} and \ref{det_Q}, one can see that this
arises because of a mismatch in the filling levels of $|0_H>$ and
$|0^\prime>$. In order to obtain a non zero overlap one would need to
insert the appropriate number of creation and annihilation operators to
balance the filling levels. In fact these operators are the t'Hooft
vertices constructed with lattice fields. Then an elegant definition
of the topological charge $q$ as seen by the fermions arises, \cite{NN1}:
\begin{equation}
q = N^- - N/2
\label{top_charge}
\end{equation}
where $q$ is naturally integer valued.

When $m_f \neq 0$ use of equations \ref{Omf}, \ref{0_H} and
\ref{Overlap_seff} yield explicit expressions for the Overlap as a
determinant of a matrix that is constructed out of $P^-$ and $Q^-$.
These expressions are used in the numerical simulation of the
Overlap ($L_s \rightarrow \infty$) in section \ref{sec-overlap-sim}.
For more details the reader is referred to \cite{NN1} and
\cite{Shamir}, \cite{Furman-Shamir}.


\section{Observables}
\label{sec-observables}
In this section the definitions of the observables that are measured
in this paper are given.  The operators involved are as in \cite{NN1},
\cite{Furman-Shamir}.

The $2n$ dimensional fermion operators of the $2n+1$ dimensional
theory are constructed out of the $2n$ dimensional fermion fields
$\psibar$, $\psi$ as in \cite{Furman-Shamir}:
\begin{eqnarray} 
\psi(x)    &=& P_R \Psi(x,0) + P_L \Psi(x, L_s-1) \nonumber \\
\psibar(x) &=& \Psibar(x,L_s-1) P_R + \Psibar(x, 0) P_L
\end{eqnarray} 
In the $L_s \rightarrow \infty$ limit of the theory these operators
exactly correspond to insertions in the Overlap of the creation and
annihilation operators discussed in section \ref{sec-model}. This will
allow explicit comparisons to be made between measurements involving
$\psibar$, $\psi$ in the $2n +1 $ theory with finite $L_s$ and
measurements with the Overlap involving the corresponding creation and
annihilation operators.

The following is a list of definitions of the observables 
for any $L_s$ and the corresponding Overlap expressions.
The definitions of the actions are as in section \ref{sec-model} but
for two flavors.  Use is made of the fact \cite{Furman-Shamir}:
\begin{equation}
\det(D_F) = \det(D_F^\dagger)
\label{det_df_dfdagger}
\end{equation} 

The fermion effective action of the $2n+1$ dimensional theory in a
background gauge field is:
\begin{equation}
e^{-S^F_{\rm eff}[L_s, m_f]} =
\int [d\Psibar d\Psi] e^{-S_F} =  
\det\left( D_F^\dagger[L_s, m_f] D_F[L_s, m_f] \right)
\label{seff_F}
\end{equation} 
The PV effective action in a background gauge field is:
\begin{equation}
e^{-S^{PV}_{\rm eff}[L_s]} =
\int [d\Psibar d\Psi] e^{-S_{PV}} =  
\det\left( D_F^\dagger[L_s, m_f=1] D_F[L_s, m_f=1] \right)^{-1}
\label{seff_PV}
\end{equation} 
The fermion effective action in a background gauge field is:
\begin{equation}
e^{-S_{\rm eff}[L_s, m_f]} = e^{-S^F_{\rm eff}[L_s, m_f] - S^{PV}_{\rm
eff}[L_s]} = { \det\left( D_F^\dagger[L_s, m_f] D_F[L_s, m_f] \right)
\over {\det\left( D_F^\dagger[L_s, m_f=1] D_F[L_s, m_f=1] \right)} }
\label{seff}
\end{equation}
\begin{equation}
e^{-S_{\rm eff}[L_s=\infty, m_f]} = <0_H| {\cal O}(m_f) |0_H>^2
\label{seff_ov}
\end{equation} 
The chiral condensate operator is:
\begin{equation}
\psibar \psi = - {1 \over 2V} \sum_x \sum_{i=1}^2
\left[ \psibar^i_R(x) \psi^i_L(x) +\psibar^i_L(x) \psi^i_R(x) \right] 
\label{pbp_op}
\end{equation}
The following observable is related to the chiral condensate in a
background gauge field:
\begin{eqnarray}
&&P_1[L_s, m_f] = 
\int [d\Psibar d\Psi]\ \psibar\psi \ e^{-S_F -S^{PV}_{\rm eff} } \nonumber \\ 
&&P_1[L_s, m_f] = - {1 \over V} \sum_x \left[ D_F^{-1}(x, L_s-1,2; x, 0,2) +
D_F^{-1}(x,0,1; x, L_s-1,1) \right] e^{-S_{\rm eff}}
\label{P1}
\end{eqnarray} 
\begin{equation}
P_1[L_s=\infty, m_f] = -{1\over V} \sum_x\left[ <0_H| \hat c_x^\dagger
{\cal O}(m_f) \hat c_x |0_H> <0_H| {\cal O}(m_f) |0_H> + (c
\rightarrow d) \right]
\label{P1_ov}
\end{equation} 
The t' Hooft vertex operator is:
\begin{equation}
w = {1 \over V} \sum_x\left[ \prod_{i=1}^2\left(
\psibar^i_R(x)\psi^i_L(x) \right) +
\prod_{i=1}^2\left( \psibar^i_L(x) \psi^i_R(x) \right) \right]
\label{w_op}
\end{equation} 
The following observable is related to the t' Hooft vertex in a
background gauge field:
\begin{eqnarray}
&&P_2[L_s, m_f] = \int [d\Psibar d\Psi] \ w \ e^{-S_F-S^{PV}_{\rm eff}} 
\nonumber \\
&&P_2[L_s, m_f] = {1 \over V} \sum_x
\left[ D_F^{-2}(x, L_s-1,2; x, 0,2) + D_F^{-2}(x,0,1; x, L_s-1,1)
\right] e^{-S_{\rm eff}}
\label{P2}
\end{eqnarray} 
\begin{equation}
P_2[L_s=\infty, m_f] = {1 \over V} \sum_x\left[ <0_H| \hat c_x^\dagger
{\cal O}(m_f) \hat c_x |0_H>^2 + (c \rightarrow d) \right]
\label{P2_ov}
\end{equation} 
The observables $e^{-S_{\rm eff}}$, $P_1$
and $P_2$ are interesting because they correspond to Overlap
expressions and, therefore, their values at the $L_s=\infty$ limit are
calculable.  Furthermore, these quantities are sensitive to the
topology of the background gauge field.
The expectation value of the fermion condensate is:
\begin{equation}
<\psibar \psi> = {1 \over Z} \int [dU] \int [d\Psibar d\Psi] \int
[d\Phi^\dagger d\Phi] \ \  \psibar \psi \ \ e^{-S} 
\label{pbp1}
\end{equation}
\begin{equation}
<\psibar \psi>_{L_s=\infty} = { \int dU {1 \over V} \sum_x\left[ <0_H|
\hat c_x^\dagger {\cal O}(m_f) \hat c_x |0_H> <0_H| {\cal O}(m_f)
|0_H> + (c \rightarrow d) \right] e^{-S_G}
\over \int dU 
<0_H| {\cal O}(m_f) |0_H>^2 e^{-S_G} }
\label{pbp1_ov}
\end{equation}
The expectation value of the t' Hooft vertex is:
\begin{equation}
<w> = {1 \over Z} \int [dU] \int [d\Psibar d\Psi] \int
[d\Phi^\dagger d\Phi] \ \ w \ \ e^{-S} 
\label{pbp2}
\end{equation}
\begin{equation}
<w>_{L_s=\infty} = { \int dU {1 \over V} \sum_x\left[ <0_H| \hat
c_x^\dagger {\cal O}(m_f) \hat c_x |0_H>^2 + (c \rightarrow d) \right]
e^{-S_G}
\over \int dU 
<0_H| {\cal O}(m_f) |0_H>^2 e^{-S_G} }
\label{pbp2_ov}
\end{equation}
Notice that a numerical evaluation of $<\psibar \psi>_{L_s=\infty}$ and
$<w>_{L_s=\infty}$ requires two separate pure gauge simulations, one
for the numerator and one for the denominator, i.e. the fermion
determinant is treated as an observable \cite{NNV}, \cite{NN2}.


\section{The free theory}
\label{sec-free}
In this section the propagator is given, its singular
part is identified and the bare fermion mass which is a function of
$L_s, m_0, m_f$, is extracted. As a verification of this result the
smallest eigenvalue of the free $2n+1$ dimensional Dirac operator is
also calculated.

The propagator has been calculated for the infinite $L_s$ case in the
first reference of \cite{NN1}. The propagator has also been calculated
for the model described in section \ref{sec-model} but only in the
limit where exponentially small contributions in $L_s$ could be
ignored \cite{Shamir}. The size of these contributions was alluded
to but no explicit expression was given.  Since in this paper the
interest is on the behavior of these contributions, the full
calculation is worked out. Because the general form of the propagator
is the same as in \cite{NN1}, \cite{Shamir} an effort has been made to
keep similar notation.

The free $2n+1$ dimensional Dirac operator of eq. \ref{D_F} in
momentum space is:
\begin{equation}
D_F(p: s,s^\prime) = i \platslash \delta(s-s^\prime) 
- b(p) \delta(s-s^\prime)
+ {1 + \gamma_5 \over 2} M(s,s^\prime) + {1 - \gamma_5 \over 2}
M^\dagger(s,s^\prime)
\label{D_F_free}
\end{equation}
and
\begin{equation}
b(p) = \sum_{\mu=1}^{2n} \left[ 1 - \cos(p_\mu) \right] + 1 -m_0
\label{bp}
\end{equation}
\begin{equation}
\plat_\mu = \sin(p_\mu), \ \ \, \mu = [1,\dots, 2n], \ \ \ 
p_\mu = {2 \pi k_\mu \over L}, \ \ \ k_\mu = [0,1, \dots L-1]
\label{plat}
\end{equation}
\begin{equation}
M(s,s^\prime) = \delta(s + 1 - s^\prime) 
- \delta(s^\prime - 0) \delta(L_s - 1 - s) (1 + m_f)
\label{RDWF}
\end{equation}
where the $\delta$ functions are understood as having ``period'' $L_s$.
Notice that the first two terms of eq. \ref{D_F_free} are the same as
for Wilson fermions with mass $(1-m_0)$, where $0<m_0<1$ (see sect.
\ref{sec-model}).  The second order free Dirac operator is diagonal in
spin:
\begin{equation}
D_F D_F^\dagger = {1 + \gamma_5 \over 2} \Omega_+ + {1 - \gamma_5
\over 2} \Omega_-
\label{D_F_D_F_free}
\end{equation}
where $\Omega_{\pm}$ have no spin indices and:
\begin{eqnarray}
&&\Omega_+(p: s, s^\prime) = [ -b(p) + M(s, s^\prime) ]
[-b(p) + M^\dagger(s, s^\prime) ] + \plat^2
\nonumber \\
&&\Omega_-(p: s, s^\prime) = \Omega_+(p: L_s -1 -s, L_s -1 -s^\prime)
\label{Omega}
\end{eqnarray}
where:
\begin{equation}
\plat^2 = \sum_{\mu=1}^{2n} \plat_\mu^2
\label{platsq}
\end{equation}

The inverse of the second order free Dirac operator $D_f D_F^\dagger$
must therefore be of the form:
\begin{equation}
G = {1 + \gamma_5 \over 2} G_+ + {1 - \gamma_5 \over 2} G_-, \ \ \ \ \
\left[ D_f D_F^\dagger \right] G = I
\label{G_free}
\end{equation}
where $G_\pm$ have no spin indices. From general considerations
$G_\pm$ must be of the form
\cite{Shamir}:
\begin{eqnarray}
G_+(p:s, s^\prime) &=& A_0 e^{- a |s-s^\prime| } + A_1 e^{-a (s +
s^\prime) } + A_2 e^{-a (L_s-1-s + L_s-1- s^\prime) } \nonumber \\ &+&
A_m\left[ e^{-a (L_s-1 + s - s^\prime) } + e^{-a (L_s-1 + s^\prime -
s) } \right] \nonumber \\ \\ G_-(p: s, s^\prime) &=& G_+(p: L_s -1 -s,
L_s -1 -s^\prime)
\label{G_pm}
\end{eqnarray}

The coefficients in eq. \ref{G_pm} are momentum dependent. The
coefficient ``a'' is the solution of the equation:
\begin{equation}
\cosh(a) = { 1 + b^2 + \plat^2 \over 2 b}
\label{a_free}
\end{equation}
A straight forward calculation results to the following expressions
for the remaining coefficients:
\begin{equation}
A_0 = { 1 \over 2 b \sinh(a) }
\label{A0_free}
\end{equation}
\begin{equation}
A_1 = - {B \over \Delta} (b - e^{-a}) (1 - m_f^2)
\label{A1_free}
\end{equation}
\begin{equation}
A_2 = { B \over \Delta} (e^a - b) (1 - m_f^2)
\label{A2_free}
\end{equation}
\begin{equation}
A_m = - { B \over \Delta} \left[ 2 b m_f \sinh(a) + e^{-a (L_s-1)}
\left(e^{-2a}\left[e^a - b\right] - m_f^2 \left[e^{-a} - b \right]
\right) \right]
\label{Am_free}
\end{equation}
where,
\begin{eqnarray}
\Delta &=& \left[ e^{2a} \left(b - e^{-a} \right) 
+ m_f^2 \left(e^a - b \right) \right] 
+ \left[ 4 m_f b \sinh(a) \right] e^{-a(L_s-1)} \nonumber \\ &+&
\left[ m_f^2 \left(b - e^{-a} \right) + e^{-2a} \left( e^a - b \right)
\right] e^{-2 a (L_s-1)}
\label{Delta_free}
\end{eqnarray}

In order to identify the singular part of the propagator an expansion
in the variables $p$, $m_f$ and $(1-m_0)^{(L_s-1)}$ is done treating
these variables as numbers with magnitudes much smaller than one. The
only singular amplitudes are $A_2$ and $A_m$ and the resulting
expression for the singular part of $G_\pm$ is:
\begin{eqnarray}
G^{\rm singular}_+(p:s, s^\prime) = 
{ 1 \over p^2 + \meff^2 } \!\!\!\!\!\!\!\!\!
&&\left\{ m_0 (2 -m_0) e^{-a (L_s-1-s + L_s-1- s^\prime) } \right.
\nonumber \\ &&\ \ \ -\left. \meff (1-m_0) \left[ e^{-a (L_s-1 + s -
s^\prime) } + e^{-a (L_s-1 + s^\prime - s) } \right] \right\}
\label{G_free_singular}
\end{eqnarray}
with,
\begin{equation}
\meff = m_0 (2 - m_0) \left[ m_f + (1-m_0)^{L_s}\right]
\label{meff_free}
\end{equation}
The expression for $G^{\rm singular}_-$ can be obtained from eq.
\ref{G_free_singular} and \ref{G_pm}. The propagator for the free
theory is then given by:
\begin{equation}
D_F^{-1} = D_F^\dagger G
\label{prop_free}
\end{equation}

To verify the above result the smallest eigenvalue of $D_F
D_F^\dagger$ is also calculated. The leading order term in an
expansion as the one used above is:
\begin{equation}
\lambda_{\rm min} = p^2 + \meff^2 + O(4)
\label{lamda_min_free}
\end{equation}
and is proportional to the inverse of the singular part of
$G_{\pm}$ as it should be.

The interesting result of the above analysis is eq. \ref{meff_free}.
This is the mass of the lightest mode of the theory and is controlled by
$m_f$, $m_0$, $L_s$. This formula strongly suggests the pattern for
chiral symmetry restoration in the model.


\section{The Interacting Theory, General Considerations}
\label{sec-interact}

The question on how the chiral symmetry restoration rate depends on
the various parameters of the model ultimately can only be answered
when the full theory is considered. This is done using numerical
simulations in section \ref{sec-hmc}. However, some general
considerations that hint to the expected behavior are useful in
order to guide the numerical experiments and to provide an
understanding of the results. Such considerations leading to specific
predictions are presented in this section.  These predictions are
confirmed by the numerical simulations and therefore sketch the
mechanisms of chiral symmetry restoration in the model.


\subsection{The Effects of Topology}
\label{sec-topology}

In this section a special kind of trajectory
\cite{Smit_Vink}, \cite{NN1} that ``cuts''
through the various topological sectors of the configuration space
is considered. The values of the relevant observables of
section \ref{sec-observables} are calculated along this trajectory.

Consider the following set of U(1) gauge link $U$ configurations
labeled by the continuous parameter $\tau \in \Re $ \cite{Smit_Vink},
\cite{NN1}:
\begin{eqnarray}
U_1(n_1, n_2) &=& \left\{ \begin{array}{ll} 1 & \mbox{if $n_1 \neq
L-1$} \\
\exp\left[ -i {2 \pi n_2 \over L} \tau \right] & \mbox{if $n_1 = L-1$}
\end{array} \right. \nonumber \\
U_2(n_1, n_2) &=& \exp\left[ i {2 \pi n_1 \over L^2} \tau \right]
\label{U1_traj}
\end{eqnarray}
where $U_1$, $U_2$ are links defined on the sites $n_1, n_2$ of the two
dimensional torus of size $L \times L$.  This configuration is
periodic along the second direction but has a discontinuity along the
first.  When $\tau$ is an integer the electric field strength is
uniform $E= 2 \pi \tau / L^2$. When $\tau$ is not an integer $E$ has a
discontinuity at $(n_1-1, n_2-1)$.  The topological charge of the gauge 
field is defined as \cite{Panag}:
\begin{equation}
q = \sum_p { \log\left[ U_p \right] \over 2 \pi i }
\label{top_charge_gauge}
\end{equation}
where the sum is over all plaquette variables $U_p$ and the logarithm
has $\log(1) = 0$ with the cut along the negative real axis. It is
straightforward to verify that $q$ is an integer that changes values
as $\tau$ is varied between integer values (see figure $1$).  For more
information on these configurations the reader is referred to \cite{Smit_Vink},
\cite{NN1}.  Here it is worthwhile to point out that this trajectory
in configuration space is interesting because it connects the trivial
configuration with uniform configurations that have non zero
topological charge.  As such, the configurations at integer $\tau$ can
be thought of as local minima and therefore in some sense they
represent vacua of different topological charge. The path that connects
the configurations of integer $\tau$ is certainly not unique but it
can nevertheless provide insightful information on how the transition
between sectors of different topological charge takes place.
Therefore, although one can not extract from this trajectory
quantitative information, the qualitative information will be quite
useful if, in particular, similar features are observed when the
full configuration space is considered in the HMC simulation of the
model in section \ref{sec-hmc}.

In the following, the observables in equations \ref{seff_F} -
\ref{pbp2_ov} are considered in the presence of the gauge field
configuration of eq. \ref{U1_traj}. The boundary conditions for both
fermion flavors in the two dimensional space are taken to be
antiperiodic. Of course, for the full dynamical $U(1)$ theory, the
choice of boundary conditions is irrelevant for as long as they are
the same for all flavors (they can change by $U(1)$ phases at the
boundary).

The $L_s = \infty$ quantities are calculated using the Overlap
formula. The finite $L_s$ quantities are obtained by explicit
computation of the determinants and inverses of $D_F$. The reader is
reminded that $e^{-S_{\rm eff}[L_s=\infty, m_f=0]}$ is not zero only
in the topological sector $q=0$. The quantity related to the chiral
condensate $P_1[L_s=\infty, m_f=0]$ is identically zero in all
sectors. This is a consequence of the exact $SU(2)_L \times SU(2)_R$
chiral symmetry. Therefore a non zero $P_1[L_s,m_f=0]$ at some finite
$L_s$ will indicate explicit breaking of the chiral symmetry. The
quantity $P_2[L_s=\infty, m_f=0]$ related to the t' Hooft vertex is
not zero only in sectors $q = \pm 1$. This represents the anomalous
breaking of the $U(1)$ axial symmetry.

In figure $1$ the topological charge along with $S_G$, $S^F_{\rm
eff}$, $S^{PV}_{\rm eff}$, $S_{\rm eff}$ and $S_{\rm eff} + S_G$ is
shown for $L=6$, $\mu l = 3.0$, $m_0 = 0.9$, $m_f=0$, and  $L_s=14$
along the trajectory of eq. \ref{U1_traj}.  The absolute scale in
these figures is of course irrelevant. One can already see the
separation of the different topological sectors with the $q=0$ sector
as the absolute minimum and the other sectors as relative minima.  It
is interesting to observe how the addition of $S^{PV}_{\rm eff}$ and
$S_G$ changes the fermion action $S^F_{\rm eff}$.

In figure $2$ $S_{\rm eff}$, $P_1$ and $P_2$ are shown along the same
trajectory and for the same parameters but for various values of
$L_s$. In the $S_{\rm eff}$ figure at $\tau=2$ the curves with larger
$S_{\rm eff}$ correspond to larger $L_s$ with $L_s=4, 6, 8, 10, 12,
14$.  The curve that sharply increases to $\infty$ at $\tau \sim 0.5$
corresponds to $L_s= \infty$.  It is clear from this figure that in
the $q=0$ sector the $L_s=\infty$ limit is approached very fast. The
$q \neq 0$ sectors carry large actions relative to the $q=0$
action and therefore they also approximate the $L_s = \infty$ limit
very well. The only regions that suffer from a slow approach to the
$L_s = \infty$ limit are the two regions in the immediate neighborhood
where $\tau$ changes from $0$ to $\pm 1$. There, the $L_s=\infty$
result raises discontinuously but the finite $L_s$ results approach
this discontinuous behavior slowly.

In the $P_1$ figure at $\tau = 0$ the curves with smaller $P_1$
correspond to larger $L_s$ with $L_s=4, 6, 8, 10, 12, 14$.  The $L_s =
\infty$ curve is identically zero for all $\tau$ and is not plotted.
From this figure it is seen that $P_1$ in the non zero topological
sectors is negligible even at $L_s=4$. In the zero topological sector
it decreases very fast with increasing $L_s$ for all values of $\tau$
except for the same two regions where $\tau$ changes from $0$ to $\pm
1$. There, two large spikes appear with heights that essentially do
not change with increasing $L_s$.  Instead, their width slowly shrinks
with increasing $L_s$. As a result, chiral symmetry is slowly
restored in these regions.

In the $P_2$ figure at $\tau = 1$, the curves with larger $P_2$
correspond to larger $L_s$ with $L_s=4, 6, 8, 10, 12, 14, \infty$.
The $L_s = \infty$ curve can be distinguished by the discontinuous behavior
at the places where $q$ changes form $0$ to $\pm1$ and from $\pm1$ to
$\pm 2$.  From this figure it is seen that $P_2$ approaches the $L_s =
\infty$ limit very fast for all $\tau$ except again for the regions
where $q$ changes.

The slow approach of $S_{\rm eff}$ and $P_2$ to the $L_s=\infty$ limit
in the regions of changing topological charge is not particularly
troubling since the effect is a small percent of the values that they
acquire along the trajectory. Not so for $P_1$. The two ``spikes''
present a large size contribution to a quantity that should otherwise
be identically zero. In figure $3$ $P_1$ is plotted vs. $L_s$ for
various ``cross sections'' of figure $2$. For $\tau = 0$ (diamonds)
$P_1$ decays exponentially. For $\tau = 0.3, 0.4$ (squares, crosses)
the rate of the exponential decay decreases. Finally, close to the point
where the configuration changes topological charge, $\tau \approx 0.55$
(octagons), there is almost no decay at all.

The slow approach to the $L_s = \infty$ limit in the neighborhoods of
changing topological charge $q$ is expected. According to the
topological charge definition eq. \ref{top_charge}, $q$ changes
because the number of negative eigenvalues of the single particle
Hamiltonian $H$ of eq. \ref{Hamiltonian} changes.  Therefore, in that
neighborhood there are configurations for which $H$ has zero
eigenvalues.  These configurations were identified in \cite{NN1} and
also discussed in \cite{Furman-Shamir}.  For these configurations the
Overlap formulation is not well defined because the ground state is
degenerate and, in turn, the finite $L_s$ theory experiences large
correlations along the $2n+1$ dimension.  However, the set of
configurations for which $H$ has an exact zero eigenvalue is of
measure zero. As a result the Overlap is a well defined formulation.
On the other hand regions of the configuration space surrounding these
special configurations are characterized by small decay rates.  The
importance of these regions is determined dynamically and only a full
simulation of the theory can accurately probe their effect to the
exponential decay. However, close to the continuum limit one would
expect that these regions are severely suppressed since the pure gauge
action separates sectors with topological charge that differs by one
unit with barriers of energy $\sim 1/g_0^2$ (see $S_G$ in figure $1$).
But away from the continuum limit, where most numerical simulations
are performed, these regions may become important and contribute to
the explicit breaking of chiral symmetry.

In order to gain some understanding about this mode of chiral symmetry
breaking the expectation value of the chiral condensate $<\psibar
\psi>$, eq. \ref{pbp1}, was calculated in a configuration space
restricted only on the trajectory of eq. \ref{U1_traj}, i.e. $<\psibar
\psi>$ was calculated as in eq. \ref{pbp1} but with the path integral
over the gauge field configuration space replaced by an integral over
the single trajectory of eq. \ref{U1_traj} parametrized by
$\tau$. This calculation was done numerically. As can be seen from
figure 2, the integrands are relatively smooth and become negligible as
$|\tau|$ is increased above $2$. Therefore, it was enough to choose a
reasonably fine discretization and sum over a finite range only. The
summation was done for step size $d\tau = 0.025$ and over the range
$|\tau| < 2.5$.  $<\psibar \psi>$ is plotted vs. $L_s$ in figure $4$
for an $L=6$ lattice, $m_0 = 0.9$, $m_f = 0$ and various gauge
couplings $\mu l / L$, $\mu l = 3.0, 2.5, 2.0 , 1.5, 1.0$ (diamonds,
squares, crosses, octagons, stars).  For $\mu l = 3.0$ there is decay
with a fast rate until $L_s=10$. For $L_s = 12 $ and above there still
is decay but with a smaller rate. As can be seen from this figure the
functional form of the fast decay is exponential.  The slower decay is
consistent with exponential but it turns out that it is also
consistent with power law or exponential times power law behavior.
Since the functional forms are further ``contaminated'' by higher
order effects in order to be able to distinguish between the different
types of decay it is estimated that calculations up to $L_s
\approx 30$ will be needed.  This is beyond the purpose of this simple
test and the available computer resources.  In any case, the important
observation is that as the gauge coupling is decreased the inflection
point moves to larger $L_s$ and the slower of the two decays becomes
faster until at $\mu l = 1.0$ there is no visible inflection below
$L_s=18$. This phenomenon can be easily understood by looking at $P_1$
in figure $2$ ($\mu l = 3.0$ there).  For $L_s$ smaller than about
$10$ the explicit breaking that occurs inside the $q=0$ sector
dominates and the contribution of the ``spikes'' to the expectation
value is small by comparison.  When this breaking has almost
completely disappeared, $L_s> 10$, the breaking that comes from the
``spikes'' dominates. This breaking disappears in a slower fashion as
the width of the ``spikes'' shrinks.  However, as the gauge coupling
is decreased the regions of changing topology where the ``spikes'' are
located are weighted less by the pure gauge action and the slower
decay is overshadowed by the initial faster exponential decay.

Therefore, one sees that there are two distinct mechanisms that
control the restoration rate of chiral symmetry. One is related to the
restoration in the $q=0$ topological sector. The other is related to
the topology changing regions of the gauge field configuration space
and in particular to the regions that connect the $q=0$ and $q=\pm 1$
sectors.

To conclude this section the effects of $m_f$ are presented.  The
observable $P_1$ along the trajectory of eq. \ref{U1_traj} is plotted
in figure $5$ for $L=6$, $\mu l = 3.0$, $m_0 = 0.9$, $m_f=0.1$
and $L_s=4, 6, 8, 10, 12, 14, 16, \infty$.  At $\tau=0$ the curves with larger
$P_1$ correspond to larger $L_s$.  It is clear that the $L_s = \infty$
limit is approached rapidly and with no complications.  Again the
region of changing topology approaches the $L_s \rightarrow \infty$ limit
slowly but it is away from it only by a small percentage.
One would expect that the non zero $m_f$ behavior will persist down to some
small value of $m_f$  before signs of the $m_f=0$ behavior described
above appear. One can visualize how figure $5$ changes with
decreasing $m_f$.  As $m_f$ becomes smaller $P_1$ tends to zero in the
various topological sectors. However, its value, at the place where the
topological charge changes from $0$ to $\pm 1$, remains roughly
constant resulting in the spikes of figure $2$ .

In order to see more clearly how the $L_s=\infty$ limit is approached for
$m_f=0.1$, the expectation value of the chiral condensate $<\psibar
\psi>$, eq. \ref{pbp1}, is calculated in a configuration space
restricted only on the trajectory of eq. \ref{U1_traj} for various
$L_s$ and for $L_s = \infty$.  In figure $6$ $\left[
<\psibar \psi>_{L_s} - <\psibar \psi>_{\infty} \right] / <\psibar
\psi>_{L_s}$ is plotted vs. $L_s$.  Again, one can see that there is an
inflection at about $L_s=10$.  However, when the inflection occurs the
$L_s = \infty$ value has already being approached to better than $0.3
\%$.  This behavior is encouraging as far as numerical simulations are
concerned.  Since most numerical simulations are done for small but
non zero masses one would expect that for some range of masses the
effects of the topology changing configurations to these simulations
will generally be small and perhaps even lost in the statistical
noise.


\subsection{The Effects of Gauge Field Fluctuations}
\label{sec-fluctuations}

In this section the effects of gauge field fluctuations to the 
chiral symmetry restoration rate are discussed.

The two mechanisms of chiral symmetry restoration identified in
section \ref{sec-topology} will be affected when the gauge field is
allowed to fluctuate. The effect of dynamical gauge fields to the
mechanism that restores chiral symmetry in the zero topological sector
can be seen by measuring $<\psibar \psi>$ in an ensemble of
configurations generated by applying small fluctuations to the trivial
configuration.  In particular, consider gauge field configurations with
links $U = e^{i r \pi}$ where $r$ is a random number in the range 
$-\epsilon < r < \epsilon$ with $\epsilon$ a small 
number that controls the size of the
fluctuations.  These configurations have a ``flat'' distribution and
in this ensemble $<\psibar \psi>$ is obtained by calculating $<P_1> /
<e^{-S_{\rm eff}}>$.  In figure $7$ $<\psibar \psi>$ is plotted vs. $L_s$ for
various values of $\epsilon = 0.4, 0.3, 0.2, 0.1$ and $0.01$
corresponding to octagons, stars, squares, crosses, and diamonds. The
value at each point was calculated in an ensemble consisting of $40$
configurations.  Antiperiodic boundary conditions have been used for
the fermions and $L=6$, $m_f=0$, $m_0=0.9$.  It can be seen that as
the size of the fluctuations decreases the chiral symmetry restoration
rate increases.

Some insight to this behavior can be gained by considering the
following ``heuristic'' argument. 
The Dirac operator of equation \ref{D_F} can be rewritten as:
\begin{eqnarray}
&&D_F = D\slash_{\rm naive} + {\cal M} \nonumber \nonumber \\
&&D\slash_{\rm naive}(x,x^\prime) = 
{1\over 2} \sum_\mu \gamma_\mu \left[ U_\mu(x) \delta(x+\hat\mu - x^\prime) -
U^\dagger_\mu(x^\prime) \delta(x^\prime+\hat\mu - x) \right] \nonumber \\
&&{\cal M} = - B + {1 + \gamma_5 \over 2} M + {1 - \gamma_5 \over 2} M^\dagger
\label{D_F_rewritten}
\end{eqnarray}
where $B$ is the Hermitian matrix given in eq. \ref{B} and $M$ is
given by equation \ref{RDWF}. For the free theory in momentum space $D_F
= i \platslash + {\cal M}(p)$ where ${\cal M}(p)$ can be read from
eq. \ref{D_F_free}. One can think of ${\cal M}(p) $ as being a
momentum dependent mass matrix.  The smallest eigenvalue of 
$\cal M M^\dagger$ is obtained at zero momentum and is equal to
$m_{\rm eff}^2$ with 
$m_{\rm eff} = m_0 (2-m_0) \left[m_f + (1-m_0)^{L_s} \right]$, 
eq. \ref{meff_free}.  The quantity $(1 - m_0)$ is the smallest 
eigenvalue of the matrix $B$.  When interactions
are turned on the smallest eigenvalue of $B$ will shift to values 
$(1-m_0^\prime)$ larger than the free theory $(1 - m_0)$ value for the simple
reason that the matrices $U_\mu(x) \delta(x + \mu - x^\prime)$ that make up
$B$ are unitary.  Therefore, the smallest eigenvalue of $\cal M
M^\dagger$ will now be ${m_{\rm eff}^\prime}^2$, $m_{\rm eff}^\prime =
m_0^\prime (2-m_0^\prime) \left[ m_f + (1-m_0^\prime)^{L_s} \right]$.
For small fluctuations one may still be able to think of $\cal M$ as a
mass matrix. If that is the case the lightest mass in the theory will
be $m_{\rm eff}^\prime$ and one sees that for $m_f=0$ the chiral
symmetry restoration rate will become faster as the fluctuations
become smaller.

Finally, the quantitative effect of dynamical gauge
fields to the mechanism related to topology changing is complicated
since it involves some understanding about the volume of configuration
space that contains gauge field configurations for which the Overlap
Hamiltonian has near zero eigenvalues.  As already
mentioned, one would expect that this volume will shrink as the continuum
limit is approached because the pure gauge action introduces energy
barriers of size $\sim 1/ g_0^2$ between sectors with topological
charge that differs by one unit.


\subsection{The range of $\bf m_0$}
\label{sec-m0_range}

In this section the allowed range of $m_0$ and the effects it has in
the approach to the continuum limit are discussed. 

In order for the doubler species to acquire masses of the order of the
cutoff $m_0$ must be in the range $0 < m_0 < 2$ \cite{Kaplan}.  
As it was found in \cite{NN1}, in order for the transfer matrix to be
positive definite for all gauge fields, the range of $m_0$ should be
further restricted to $0<m_0<1$, eq. \ref{m0_range}.  For QCD, any
value of $m_0$ in this range should lead to the same continuum limit
since the local and global symmetries of the theory remain unchanged.
On the other hand, different choices of $m_0$, will result in different
ways of approaching this limit. For the Schwinger model there is an
additional complication. As mentioned in the introduction, a
four-Fermi term is a marginal operator in two dimensions.  Although it
is not explicitly introduced in the action, the DWF regularization
introduces such a term with a coefficient that depends on $m_0$
\cite{NNV}, \cite{NN2}.  Therefore, different values of $m_0$ will
lead to different continuum limits. For this reason, a quantitative
study of the effects of $m_0$ to the approach to the continuum limit
in the Schwinger model is complicated and will not be done here.
However, there are some important generic features that can be
discussed.

Consider the Overlap Hamiltonian $H$, eq. \ref{Hamiltonian} as a
function of the $2n+1$ dimensional mass $m_0$ \cite{NN1}.  It is easy
to see that for $m_0 = -\infty$, $H$ has the same number of positive
and negative eigenvalues, $N_+ = N_-$.  It is also easy to see that
for $m_0 < 0$, $H$ can not have a zero eigenvalue. As a result $H$ has
$q = 0$, for $m_0 < 0$.  As $m_0$ is increased from zero, $H$ can
develop zero eigenvalues and as a result $q \neq 0$ (eigenvalues cross
zero altering $N^-$).  Close to the continuum it can be easily seen
that most crossings will occur around $m_0 = 0$ \cite{NN1}.  Farther
away from the continuum pure gauge numerical simulations
\cite{NV} show a finite region $[m_{0_{\rm min}}, m_{0_{\rm max}}]$ where
most crossings occur.  This suggests that one should keep 
$m_{0_{\rm max}} \ll m_0$ for two reasons.  First, 
one would like to be as far away as possible from the region where
crossings occur since, as discussed in section \ref{sec-topology},
it is there that the decay rates become
small. Second, as can be seen from the definition of the topological
charge, eq. \ref{top_charge}, it is only then that the full effects of
non-trivial topology are visible to the fermions.  In particular, one
would expect that by keeping $m_{0_{\rm max}} < m_0$ quantities that are
sensitive to topology will have a ``smoother'' approach to the
continuum limit. 

From the above discussion it appears that at any coupling the safest
choice would be to set $m_0$ to its largest allowed value $m_0
\approx 1$. Even then, because in general $0<m_{0_{\rm min}} 
< m_{0_{\rm max}} < 2$ it may be that far away from the continuum 
$1 < m_{0_{\rm max}} \Rightarrow m_0< m_{0_{\rm max}}$.  
If this situation occurs the
finite lattice spacing errors at that coupling will be large and the
chiral symmetry restoration rate will be slow.  However, the
numerical simulations of \cite{NV} indicate that even for reasonably
strong couplings $m_{0_{\rm max}} < 1$.  The couplings used in this
paper fall in this category (see figure $1$ in \cite{NV}) and $m_0$ is
kept fixed to $m_0 = 0.9$.

In \cite{Blum-Soni} it was found that for QCD some tuning ($m_0 =
1.7$) was necessary in order to sufficiently restore chiral symmetry
at $L_s=10$.  However, the configurations used there were not
generated with the DWF action but rather with the staggered fermion
action at $\beta = 5.7$.  It is possible that this large value of
$m_0$ is at least due in part to the ``semi-quenched'' nature of the
calculation and if the configurations are generated with the DWF
action setting $m_0 \approx 1$ may be sufficient.


\section{Dynamical Simulation of the Massive Theory with the Overlap} 
\label{sec-overlap-sim}

In this section a full dynamical simulation that measures the chiral
condensate $<\psibar \psi>$ and the t' Hooft vertex $<w>$ of the two
flavor massive Schwinger model using the Overlap formalism of the
model of section \ref{sec-model} will be presented. The result is
interesting on its own right and it also provides the $L_s=\infty$
numbers that will be used to compare with the results of the dynamical
simulation for finite $L_s$ presented in the next section.

The expectation values $<\psibar \psi>$ and $<w>$ can be calculated
numerically using equations \ref{pbp1_ov} and \ref{pbp2_ov}
respectively. As it is evident from these equations a pure gauge
simulation with action $S_G$ must be performed with the overlap
factors appearing in the numerator and denominator treated as
observables.  The pure gauge theory expectation values of the overlap
factors in the numerator and denominator are then divided to produce
$<\psibar \psi>$ or $<w>$. A heat bath algorithm was used to generate
configurations with the standard Wilson plaquette pure gauge action of
eq. \ref{action_G}.

Measurements of $<w>$ for the massless, $m_f=0$, theory were performed
in \cite{NNV}. The reader is referred there for more details on the
method and results. However, the result for $w$ in that reference can
not be compared directly with the $m_f=0$ result here since a
different implementation of the Overlap was used. Furthermore, the
single plaquette action used there was a heat kernel action while the
standard Wilson plaquette action is used here.  The result for $<w> /
m_\gamma^2$ vs. $L_s$ on an $L=6$, $m_0 =0.9$, $\mu l = 3.0$ lattice
is given in figure $8$.  This quantity acquires a non zero vacuum
expectation value even for $m_f = 0$ (dotted lines) as a result of the
anomalously broken $U(1)$ axial symmetry.

The result for $<\psibar \psi> / m_\gamma$ vs. $m_f$ on the same
lattice and for the same parameters is presented in figure $9$.  The
behavior of $<\psibar \psi>$ vs. $m_f$ is interesting in that
$<\psibar \psi> \sim m_f$ for $m_f < 0.1$ while $<\psibar \psi> \sim
m_f^{1/3}$ for $m_f > 0.1$. Fits to $A m_f^p$ for $m_f < 0.1$ and for
$m_f > 0.1$ are shown in the same figure. Both fits have a $\chi^2$
per degree of freedom of about one.  For $m_f < 0.1$ $p = 0.996(3)$
while for $m_f> 0.1$ $p = 0.32 (2)$.

This type of behavior was found by analytical continuum calculations in
\cite{Smilga-Hetrick}. In particular, the linear behavior was found to take place
for $m_f L \ll 1$ while the $m_f^{1/3}$ behavior was found to occur
for $m_f L \gg 1$. Unfortunately, the coefficients calculated in that
reference can not be directly compared with the ones here because, as
previously discussed, an additional four-Fermi interaction is
induced by the DWF regularization \cite{NNV}, \cite{NN2}. This term
in two dimensions is not irrelevant and it will contribute to the
continuum limit. For the same reasons $<w>$ can not be
directly compared with the continuum results of \cite{Smilga-Hetrick}.


\section{Hybrid Monte Carlo Simulation}
\label{sec-hmc}

In this section a full dynamical simulation of the two flavor
Schwinger model for finite $L_s$ is presented.  The algorithm used is
the standard Hybrid Monte Carlo (HMC) algorithm \cite{DKPR}.  The
expectation value of the chiral condensate $<\psibar \psi>$, eq.
\ref{pbp1}, is calculated and used to monitor the amount of explicit
chiral symmetry breaking. The reason for using $<\psibar \psi>$
instead of the pion mass is that in two dimensions there is no
spontaneous chiral symmetry breaking and although the pion mass
vanishes for vanishing fermion mass the pion is not a Goldstone
particle. In this sense $<\psibar \psi>$ is as good a probe of chiral
symmetry breaking as is the pion mass, but unlike the pion mass it has
the practical advantage of not requiring large lattice size along the
time direction in order to measure the decay of the pion correlator.
The effects of the anomalously broken $U(1)$ axial symmetry are
monitored by measuring the expectation value of the t' Hooft vertex
$<w>$, eq. \ref{pbp2}.

The smaller size lattices were simulated on the workstations of the
Physics Department of Columbia University. The larger lattices were
simulated on the Silicon Graphics Power Challenge Array computer
system at NCSA UIUC and also on the C90 supercomputer at PSC.


\subsection{The algorithm}
\label{sec-hmc-algorithm}

The HMC Hamiltonian is given by:
\begin{eqnarray}
H_{\rm HMC} &=& {1 \over 2} P^2 + S_G[U] 
+ \chi^\dagger \left( D_F^\dagger [m_f, L_s] D_F[m_f, L_s] \right)^{-1} \chi
\nonumber \\
&+& \Phi^\dagger \left( D_F^\dagger [m_f=1, L_s] D_F[m_f=1,L_s] \right) \Phi
\label{hmc_hamiltonian}
\end{eqnarray}
where $P$ is the HMC momentum, $S_G$ is the pure gauge plaquette
action given in eq. \ref{action_G}, $U$ is the $U(1)$ gauge field,
$\chi$ is the pseudofermion field, $\Phi$ is the bosonic PV field and
$D_F$ is the three dimensional Dirac operator of eq. \ref{D_F}.

The HMC trajectory length $\tau$ is set to $\tau = 1$, the step size
is $\tau / N$ and the number of steps $N$ is adjusted
according to the size of the effective bare fermion mass. Typical simulations
with $m_f \ge 0.05$ are done with $50$ HMC steps. For the $m_f=0$ simulations
the number of steps is set to $100-400$ depending on the
size of $L_s$. This is necessary because, for $m_f=0$, trajectories that
cross between different topological sectors experience large HMC fermion
forces (for more details see section \ref{sec-hmc-topology}).
The Conjugate Gradient residual is set to $10^{-8}$. Typically,
the number of Conjugate Gradient iterations is around $50$ and does
not exceed $\sim 100$ for $m_f=0$ and $L_s=14$. The number of
measurements is $\sim 8,000$ except for the more ``expensive''
$12 \times 12$, $10 < L_s$ lattices where $1,000 - 2,000$
measurements were performed.


\subsection{HMC and topology}
\label{sec-hmc-topology}

In this section problems related to the sampling of non zero topological
sectors with the HMC algorithm at small fermion masses are discussed.

The HMC algorithm is very successful provided the fermion mass does
not become very small. If the fermion mass becomes very small then the
effects of topology are not reproduced correctly.  In particular, the
measurement of $<w>$ becomes problematic.  If, for some fixed
volume, the fermion mass is made very small then
similar analysis as in \cite{LS} indicates that the effect of the zero
modes coming from different topological sectors becomes important. In
particular, as can be seen from the Overlap implementation \cite{NN1},
\cite{NNV}, at finite physical volume, at zero mass and in the $L_s
\rightarrow \infty$ limit, the operator $w$ receives contributions only
from sectors $\pm 1$, while the fermion Boltzman weight (fermion
determinant) is not zero only in sector 0.  As $m_f$ is turned on
(and/or $L_s$ is decreased from infinity) the fermionic determinant
becomes non zero in sectors other than $0$ and the operator $w$
receives contributions from sectors other than $\pm1$.  
Because of this behavior at
small fermion mass the HMC algorithm will mostly sample the sector $0$ where
the observable $w$ receives small contributions. The algorithm will
infrequently visit the sectors $\pm 1$, but when it does the observable
$w$ will receive large contributions to ``make up'' for the small sampling
rate. As a result, when the fermion mass is decreased a larger number of HMC
iterations will be needed to sample the $\pm 1$ sectors correctly.
Therefore, for a fixed amount of computer time, if the fermion mass
becomes very small the important contributions may not even be sampled
at all and as a result not only the expectation value of $w$ will be
underestimated but also the associated error.  This type of difficulty
has already been noticed in simulations of the Schwinger model
\cite{hmc-difficulties}, \cite{PMV}.

The problem described above leaves a clear signature in the time history
of $w$. In figure $10$ the time history of $w$ is given for six
different fermion masses at $L_s=14$. As the fermion mass is decreased, the
average around which  $w$ fluctuates decreases. This decrease
is compensated by the large contributions received from the 
$q=\pm 1$ sectors. These contributions start to appear as ``spikes'' in the
time history. The smaller the fermion mass the larger the ``height''
of the spikes but since the corresponding Boltzman weight becomes
smaller their frequency also decreases. Notice the different
scale of the $m_f=0.05$ and $m_f=0.0$ graphs.

But it is not only $<w>$ that is affected at small fermion masses. If,
along an HMC trajectory, the topological charge changes then the
fermion determinant changes by a large amount. As a result, the HMC
fermion force becomes large and the HMC step size errors become
large. Therefore, as the fermion mass is made smaller one must
adjust the HMC step size accordingly.
Finally, it should be stressed that these difficulties are not
related to DWF but rather to the HMC algorithm. In particular, the
simulations using the overlap in section \ref{sec-overlap-sim} do not
suffer from these problems since there the fermion determinant is
treated as an observable in a pure gauge heat bath.


\subsection{$\bf m_f = 0$}
\label{sec-hmc-mf-zero}

In this section a strict test of chiral symmetry restoration is
done. 

A full dynamical simulation using the HMC algorithm is performed with
the explicit fermion mass $m_f$ set to zero so that the only breaking
of chiral symmetry comes from the finite extent $L_s$. The restoration
rate at fixed physical volume and various lattice spacings is studied
by measuring $<\psibar \psi>$ for various values of $L_s$.  Although
this provides a strict test one must keep in mind that the non-zero
topological sectors may be suppressed more than they should for the
reasons mentioned in section \ref{sec-hmc-topology}. This means that
the rate of restoration of chiral symmetry observed here is mainly due
to effects that occur in the zero topological sector and its vicinity.
The following results were obtained at fixed $m_0=0.9$.

In figure $11$, $<\psibar \psi> / m_\gamma$ is plotted in a ``log''
plot vs. $L_s$ at fixed physical volume $\mu l = 3.0$ and for various
lattice spacings $\mu l / L = \mu a$ where $L=6, 8, 10, 12$
corresponding to the lines from top to bottom.  Data for $L=4$, $L_s =
6 - 10$ are statistically indistinguishable from the $L = 6$ data and
are not plotted.  For $L_s = 6 - 10$ the decay is consistent with
exponential with a rate that becomes faster as the lattice spacing
decreases. For $L_s = 12-22$ the decay is again consistent with
exponential but with a slower rate. Again, this rate becomes faster as
the lattice spacing decreases.  Also, the percent change of the rate
at $L=6$ is $\approx 54 \pm 6\%$ but at $L=12$ is $\approx 31 \pm
6\%$.  The fits shown are two parameter fits to $<\psibar \psi> /
m_\gamma = B e^{ -c L_s}$.  The $\chi^2$ per degree of freedom is
smaller than one for all the fits except for the $L=12$, $L_s=6-10$
data that have a $\chi^2$ per degree of freedom of $\approx 3$.

The exponentiated rate $e^{-c}$ of the various fits vs. $1/L \sim a$
is shown in figure $12$.  The diamonds correspond to the $L_s=6 - 10$
fits while the crosses to the $L_s = 10 - 22$ fits.  One can see that
$e^{-c}$ is roughly a linear function of $1/L \sim a$ for the $L_s =
6-10$ fits and for $L=8, 10, 12$.  However, more data at smaller
lattice spacings are needed before one can be confident that scaling
has set in and that this is the correct scaling form.

Although the above fits are all consistent with exponential decay,
power law decay of the form $<\psibar \psi> / m_\gamma = B {L_s}^{-p}$
can be excluded with some confidence only for the fast decay, $L_s=6-10$,
at the smallest lattice spacing, $L=12$. A power law fit
to this data has $\chi^2$ per degree of freedom of $\approx 32$.  More
statistics and larger $L_s$ will be needed in order to clearly
establish the type of decay for the other cases. For example, at the
largest lattice spacing, $L=6$, it is estimated that the error bars
will have to be reduced from their few percent size down by a factor
of about ten. Alternatively, the error bars can be kept at the few
percent level but then it is estimated that $L_s$ will have to be made
as large as $\approx 30$.  Both approaches are beyond the purpose of
this paper and the available computer resources.

The facts that the decay changes for $L_s > 10$, that the slower decay
approaches the faster one as the continuum limit is approached and
that both decays become faster as the continuum limit is
approached can all be understood from the analysis in sections
\ref{sec-topology} and \ref{sec-fluctuations}.
Finally, the t' Hooft vertex $<w>$ was also measured but, as expected
from the discussion in section \ref{sec-hmc-topology}, its value and
associated error is underestimated. In particular, its value is much
lower than the corresponding Overlap value of section
\ref{sec-overlap-sim}.  For this reason that data is uninteresting and
is not presented here.


\subsection{$\bf m_f \neq 0$}
\label{sec-hmc-mf}
In this section the $m_f \neq 0$ case is studied. Since
typical QCD simulations are done for non zero fermion mass
the results of this section are of practical interest.

Dynamical simulations are performed with masses $m_f$ large enough,
$0.1 \le m_f$, so that the effects of topological sectors $q=0, \pm 1$
are not miscalculated due to problems associated with the HMC
algorithm as described in section \ref{sec-hmc-topology}.  Both
$<\psibar \psi>$ and $<w>$ are measured and their approach to the $L_s =
\infty$ limit is studied and compared with the $L_s=\infty$ results of
section \ref{sec-overlap-sim}. This is done for fixed physical volume,
and various lattice spacings. The parameter $m_0$ is kept fixed at
$0.9$.

In figure $13$ $\left[ <\psibar \psi> / m_\gamma \right]^3$ is plotted
vs. $L_s$ for $m_f = 0.1, 0,2, 0.3, 0.5$ at fixed physical volume and
lattice spacing, $\mu l = 3.0$, $L=6$.  According to the results in
section \ref{sec-overlap-sim} one expects that $\left[ <\psibar \psi>
/ m_\gamma \right]^3 \sim m_{\rm eff}$. Therefore a fit of $\left[
<\psibar \psi> / m_\gamma \right]^3$ vs. $L_s$ is made to the form $A
+ B e^{-c L_s}$. All fits have a $\chi^2$ per degree of freedom
$\approx 1-2$.  In these figures the cross is the coefficient A and
the dotted lines are the $L_s = \infty $ result of figure $9$
plus/minus the error. One can see that as $m_f$ becomes larger the
$L_s = \infty $ result is approached faster.  This is in accordance with
naive expectations born out from the free theory formula for $m_{\rm
eff}$ eq. \ref{meff_free}.  In figure $14$ $<w> / m_\gamma^2$ is
plotted vs. $L_s$ for the same parameters as in figure $13$. The fits
are again to a form $A + B e^{-c L_s}$ and have $\chi^2$ per degree of
freedom $\approx 1-2$. One can see that the $L_s=\infty$ result has
already been approached at $L_s=6$.

The effects of changing the lattice spacing at $m_f=0.2$ can be seen
in figures $15$ and $16$. In figure $15$ $\left[ <\psibar \psi> /
m_\gamma \right]^3$ is plotted vs. $L_s$ at fixed physical volume $\mu
l = 3.0$ for different lattice spacings $\mu l / L = \mu a$, $L=4, 6,
8, 10$.  The fits are again to a form $A + B e^{-c L_s}$ and have a
$\chi^2$ per degree of freedom of $\approx 1-2$.  One can see a
similar behavior as the one in section \ref{sec-hmc-mf-zero}.  As the
lattice spacing is reduced the rate $c$ of the exponential approach to
the $L_s=\infty$ result increases.  For example, $<\psibar \psi>$ at
the larger lattice spacing $L=4$ decays with $c=0.54(3)$ but at the
smaller lattice spacing $L=10$ it decays faster with $c=1.1(1)$.
However, one should note that for $L=8$ and $L=10$ the rate saturates
and is essentially dictated by the $L_s=4, 6$ points with the $L_s=6$
point very close to the $L_s=\infty$ result.  If a second slower rate
sets in for $10 \ltapprox L_s$ it is unimportant and is lost in the
statistical noise.  Some insight to this behavior can be gained from
the analysis at the end of section \ref{sec-topology} (in particular
see figure $6$).  Similar behavior is observed in figure $16$ for $<w>
/ m_\gamma^2$. All fits have $\chi^2$ per degree of freedom $\approx
1-2$.

Similar results are obtained if one keeps the physical volume
and $m_f$ in physical units fixed while changing the lattice spacing.
This can be seen in figures $17$ and $18$ by comparing the $L=10$,
$m_f=0.2$ data (diamonds) with data at $L=4$, $m_f=0.5$
(squares). In these graphs the physical volume is fixed at $\mu l =
3.0$, and $m_f$ in physical units is fixed at $m_f L = 2.0$.
Again the decay rate increases as
the lattice spacing is reduced.  $<\psibar \psi>$ at the larger
lattice spacing $L=4$ decays with $c=0.48(5)$ but at the smaller
lattice spacing $L=10$ decays faster with $c=1.1(1)$. All fits
have $\chi^2$ per degree of freedom $\approx 1-2$.

Finally, it should be noted that although the above data is consistent
with exponential decay, other types of decay can not be excluded. This
is mainly due to the fact that since the $L_s=\infty$ result has
already being approached, within statistics, for $L_s=8-10$, the decay is
basically dictated only by the two points $L_s=4,6$. More statistics
are needed in order to clearly establish the type of decay.


\section{Conclusions}
\label{sec-conclusions}

In this paper the properties of Domain Wall Fermions (DWF) were
studied in the context of the two flavor Lattice Schwinger model.  The
expectation value of the chiral condensate $<\psibar \psi>$ was used
to probe issues related to restoration of the regularization induced
chiral symmetry breaking. The expectation value of the relevant t'
Hooft vertex $<w>$ was used to probe issues related to topology.

Dynamical numerical simulations of the full theory were performed. It
was found that, as expected from perturbative considerations, the
restoration of chiral symmetry as a function of $L_s$ ($L_s$ is the
size in lattice units of the $2n+1$ direction) at a fixed physical
volume and lattice spacing, is consistent with exponential decay.  In
particular, the data is consistent with a picture where chiral
symmetry is restored with a fast exponential decay rate for $L_s$ up
to some value and with a slower exponential decay rate for $L_s$ above
that value.  For the range of lattice spacings used in this paper the
inflection appeared at $L_s \approx 10$.  Using a simple model it was
found that the first fast decay can be associated with restoration of
chiral symmetry in the zero topological sector while the second slower
decay can be associated with the regions of gauge field configuration
space that connect the $q=0$ and $q=\pm 1$ topological sectors.

The effects of the size of the lattice spacing $a$ to the two decays
were studied using both analytical arguments and explicit numerical
simulations of the full theory. It was found that for zero explicit
fermion mass the fast decay associated with the zero topological
sector becomes faster as the lattice spacing is decreased.  The
vanishing of the chiral condensate is consistent with a form $e^{-c
L_s}$ with $e^{-c}$ being roughly a linear function of $a$, but more
data at smaller lattice spacings are needed before one can be
confident that scaling has set in and that this is the correct scaling
form.  The second slower decay also becomes faster as the lattice
spacing is decreased and it differs less from the faster decay as the
lattice spacing becomes smaller. For the smallest lattice spacing
studied the two exponential decay rates differed by 
$\approx 31 \pm 2 \%$.

For small but non zero explicit fermion mass $m_f$ the values of
$<\psibar \psi>$ and $<w>$ were measured. The corresponding $L_s =
\infty$ numbers were calculated by performing numerical simulations
with the Overlap formalism.  It was found that the $L_s = \infty$
numbers were also approached in a way that is consistent with
exponential decay with a rate that became faster as the lattice
spacing decreased. Furthermore, the larger the fermion mass the sooner
the $L_s = \infty$ value was approached and for the fermion masses
studied in this paper the $L_s = \infty$ result was already achieved
to within a few percent for $L_s = 4 - 8$.  If a second slower decay
does set in for $10 \ltapprox L_s$, it is unimportant and was lost in
the statistical noise.  Finally, an interesting result was obtained
from the measurements of $<\psibar \psi>$ and $<w>$ vs. $m_f$. It was
found that these measurements are in agreement with the analytical
predictions of \cite{Smilga-Hetrick}. In particular the interesting
$<\psibar \psi> \sim m_f^{1/3}$ behavior was reproduced.

Although all the numerical data are consistent with exponential decay,
power law decay can be excluded only for the fast decay at the
smallest lattice spacing studied. For that data the decay is
sufficiently fast and the error bars are sufficiently small so that a
power law fit can be safely excluded since it has a $\chi^2$ per degree
of freedom $\approx 31$. More statistics and larger $L_s$ are
needed in order to be able to clearly distinguish between exponential
and power law decay for the rest of the data points.

The next step is to carry out a similar investigation for dynamical
QCD. Many of the characteristics of DWF found here are sufficiently
generic so that one would expect that they will also be present in
QCD. If it turns out that QCD at the presently accessible lattice
spacings, volumes and quark masses has similar restoration rates as
the ones found here, then DWF will indeed provide a powerful fermion
discretization method.


\section*{Acknowledgments}

The author would like to thank N. Christ, R. Mawhinney, R. Narayanan
and H. Neuberger for many useful discussions. This research was
supported in part by the DOE under grant
\# DE-FG02-92ER40699. This work was also  partially supported by the National Center for
Supercomputing Applications under grant \# PHY970002N and utilized the
Silicon Graphics Power Challenge Array computer system at the National
Center for Supercomputing Applications, University of Illinois at
Urbana-Champaign. Also, this research was supported in part by grant
\# PHY960005P from the Pittsburgh Supercomputing Center and
utilized the C90 supercomputer.


\vfill


\if \epsfpreprint Y

\eject

\figure{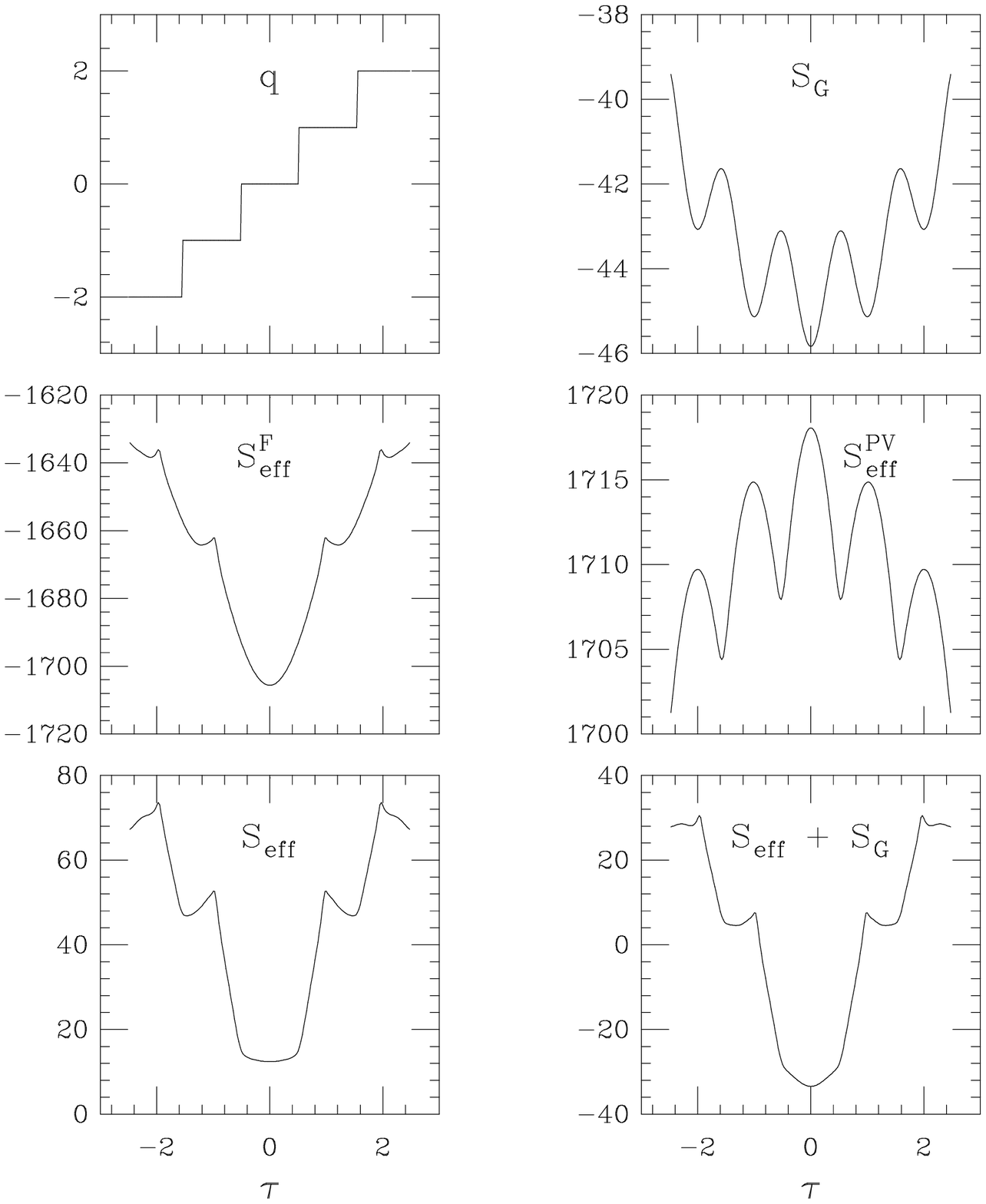}{\one}{\figsizeC}

\figure{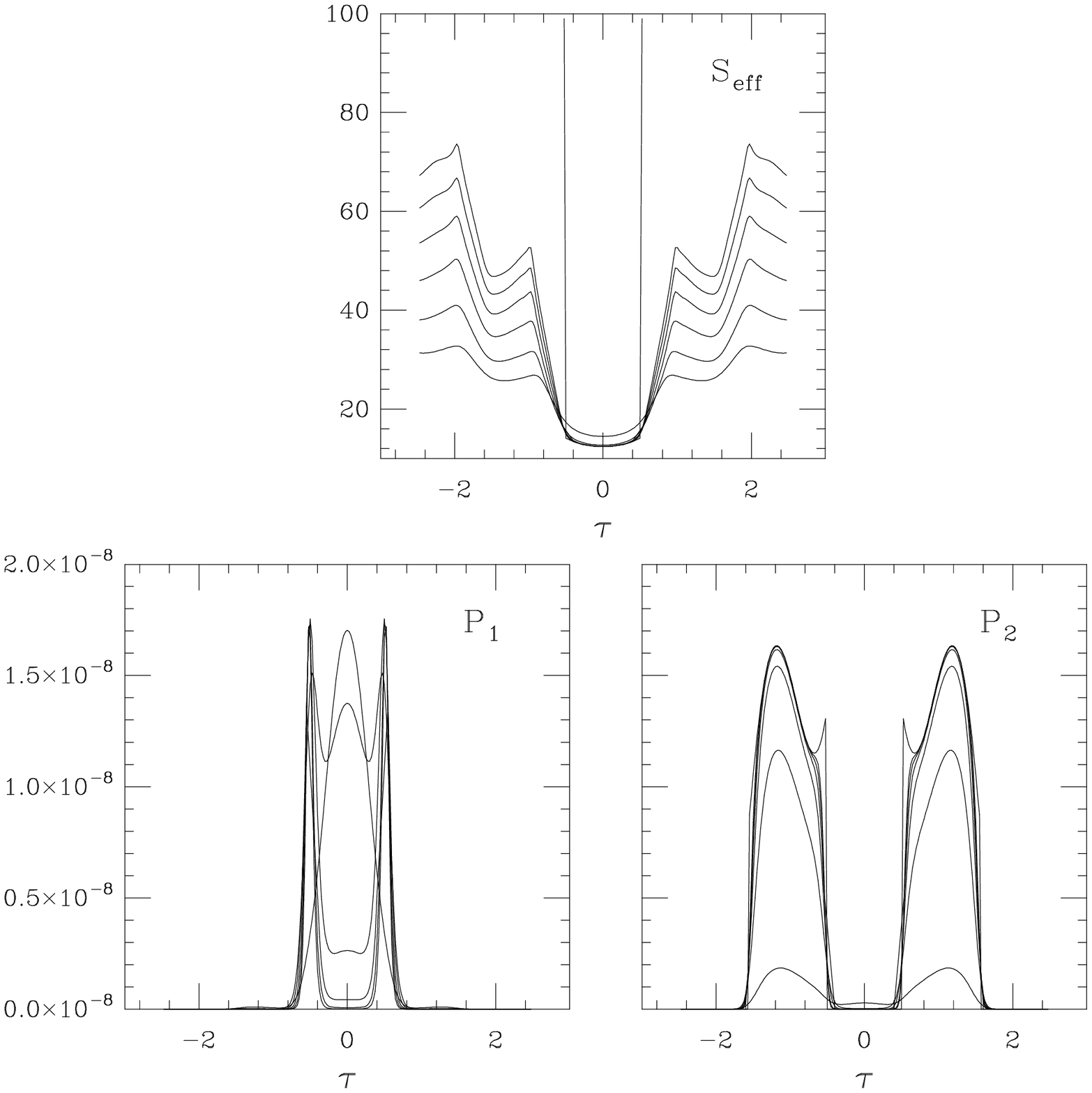}{\two}{\figsizeA}

\figure{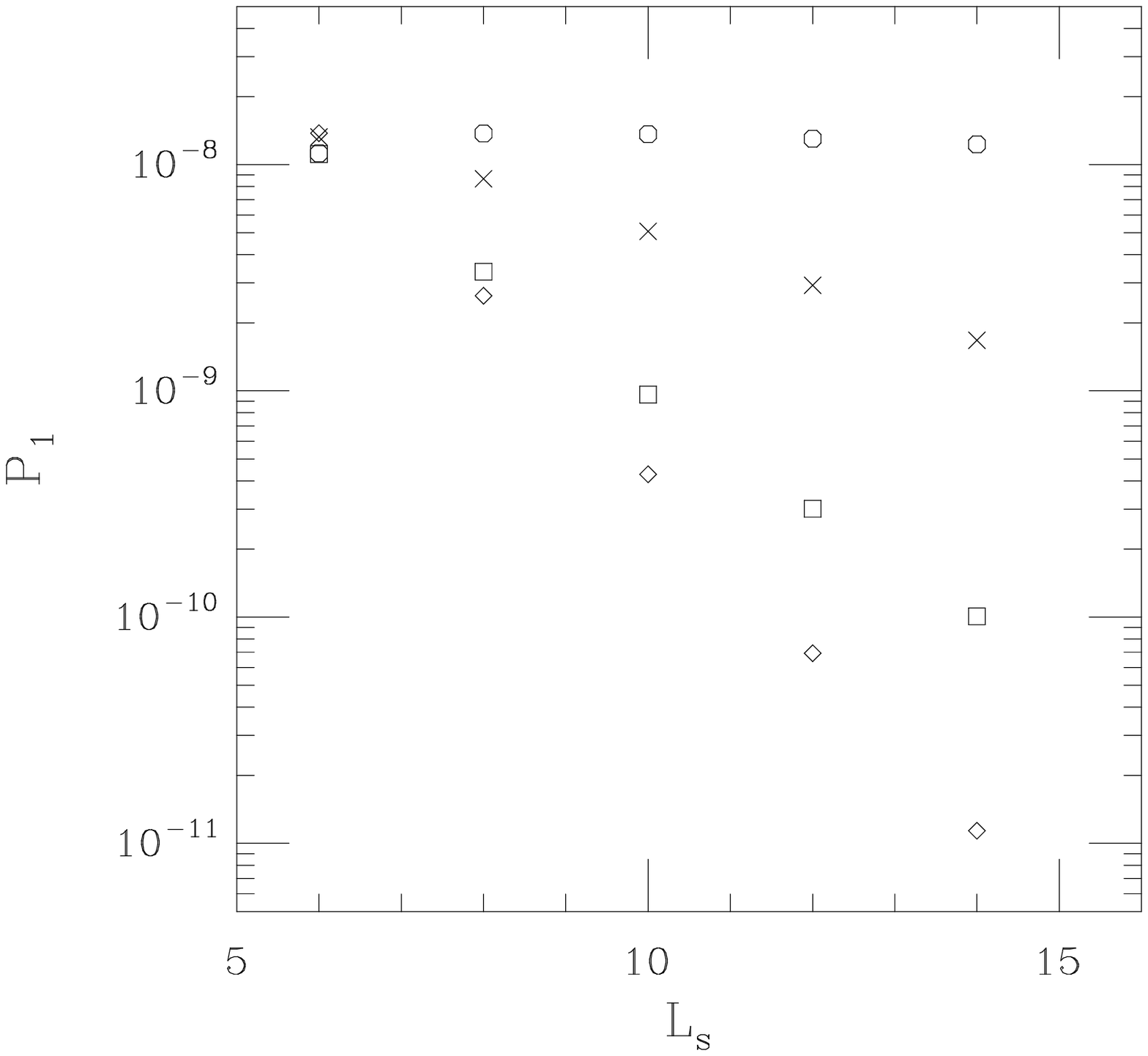}{\three}{\figsizeB}

\vskip -0.7 truein

\figure{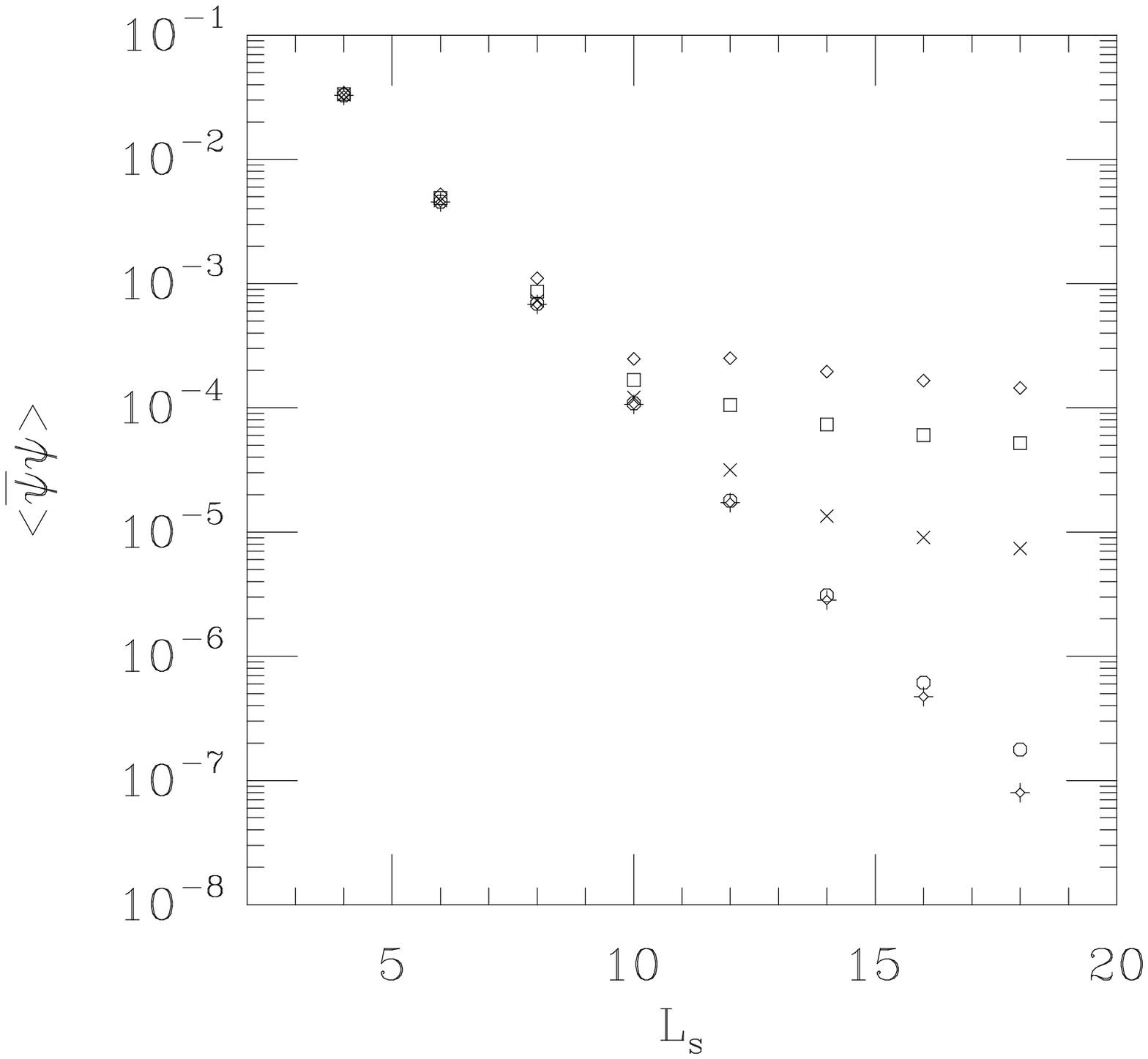}{\four}{\figsizeB}

\figure{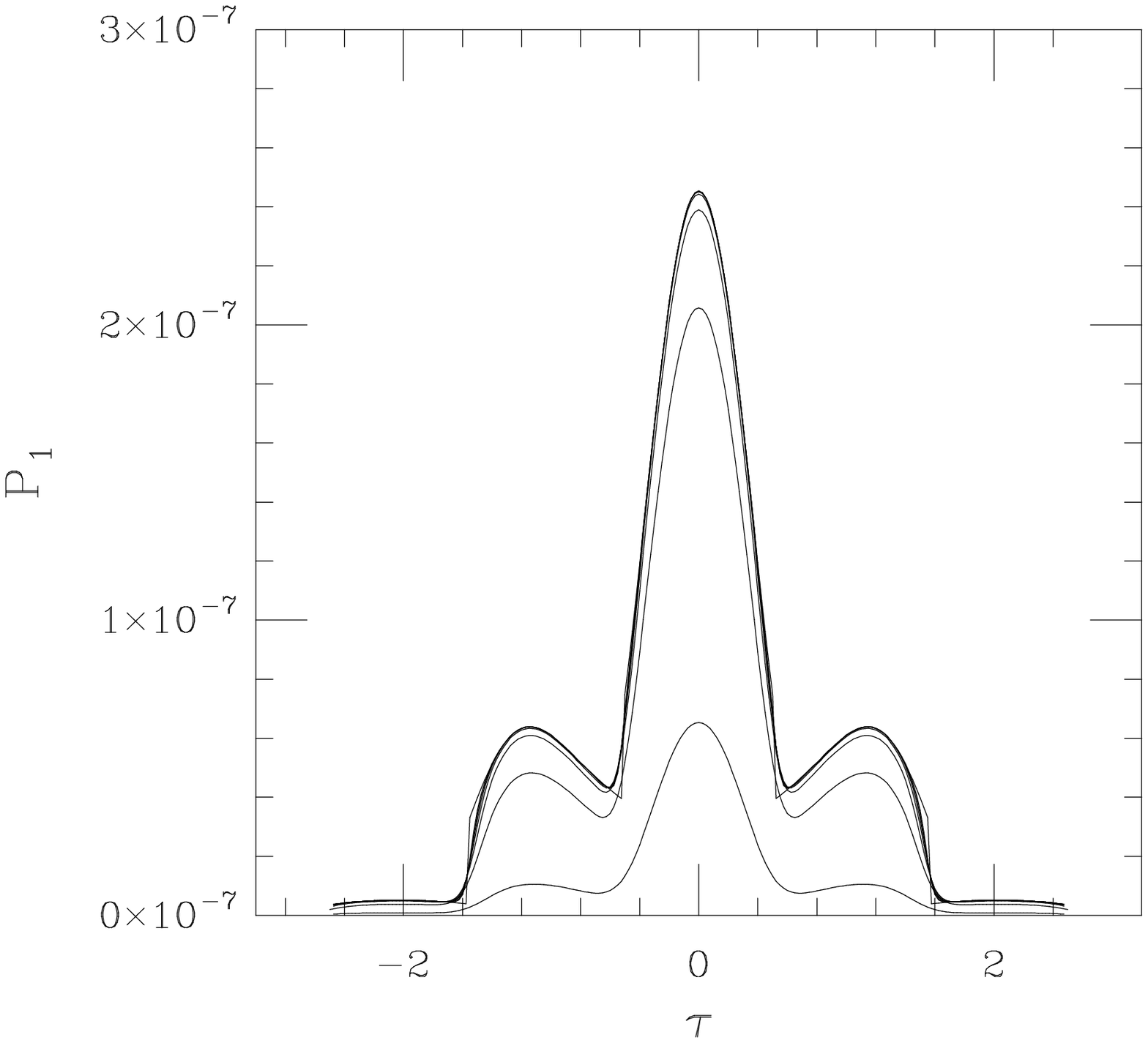}{\five}{\figsizeB}

\vskip -0.7 truein

\figure{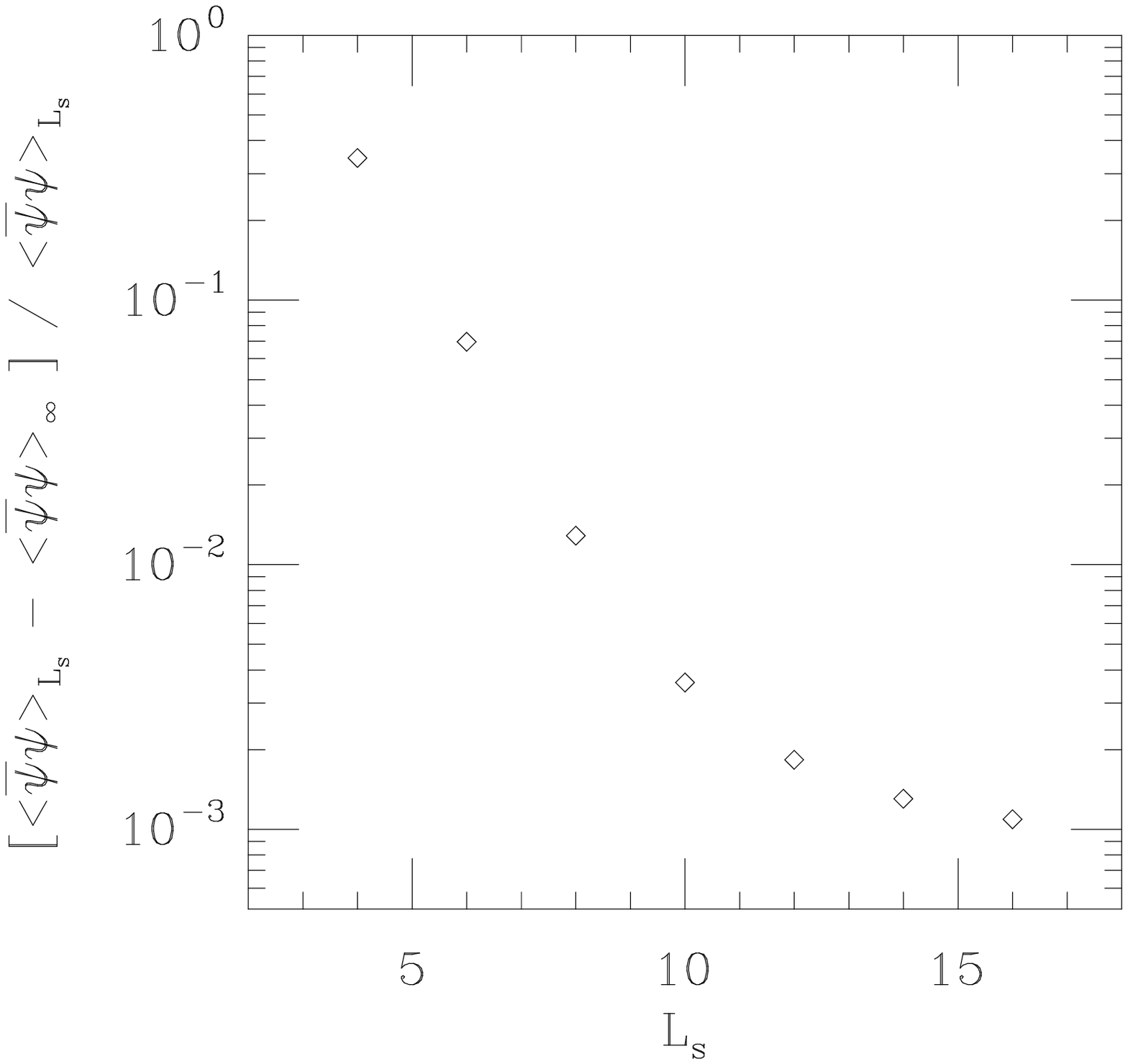}{\six}{\figsizeB}

\figure{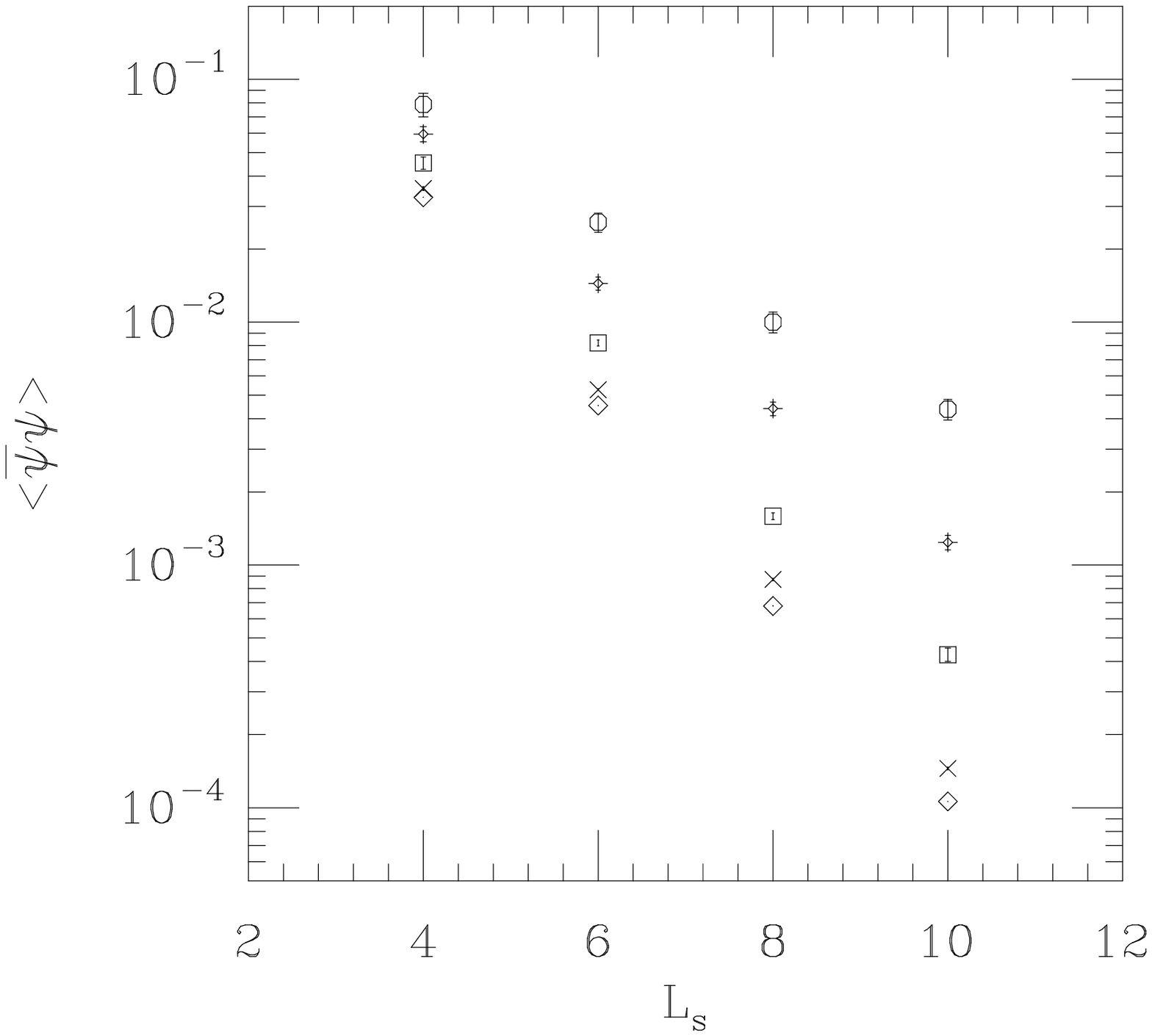}{\seven}{\figsizeB}

\vskip -0.7 truein

\figure{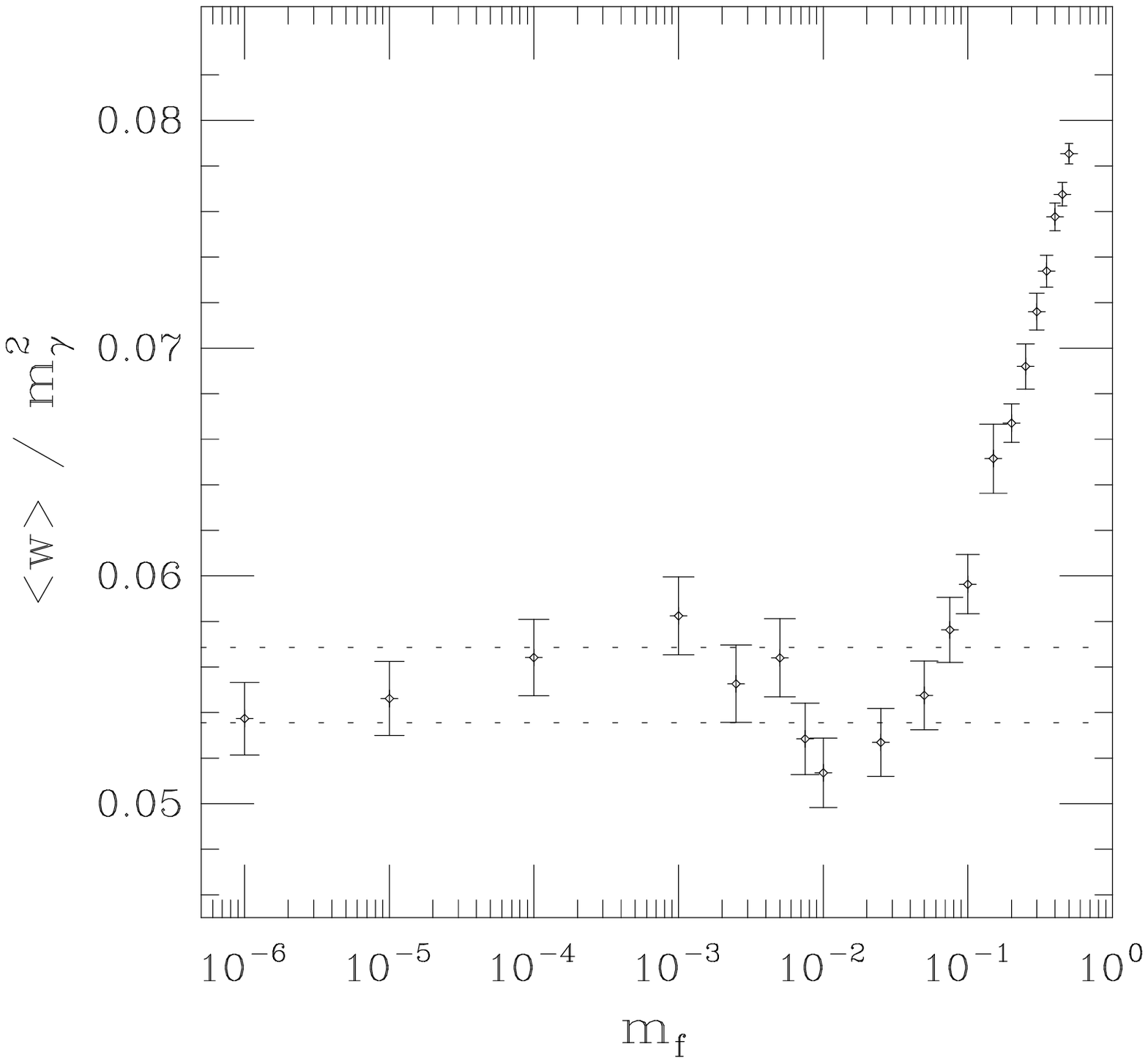}{\eight}{\figsizeB}

\figure{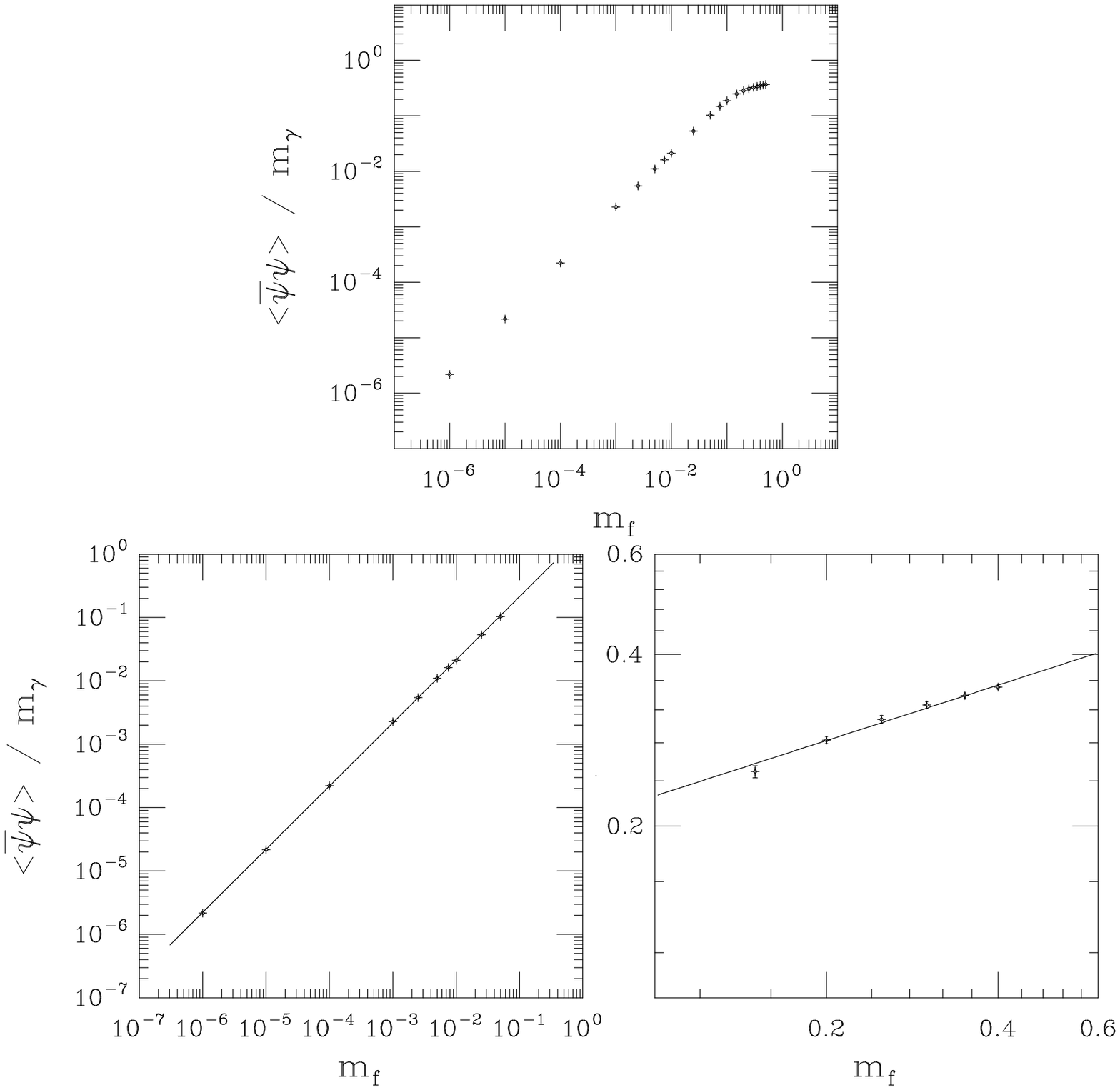}{\nine}{\figsizeA}

\figure{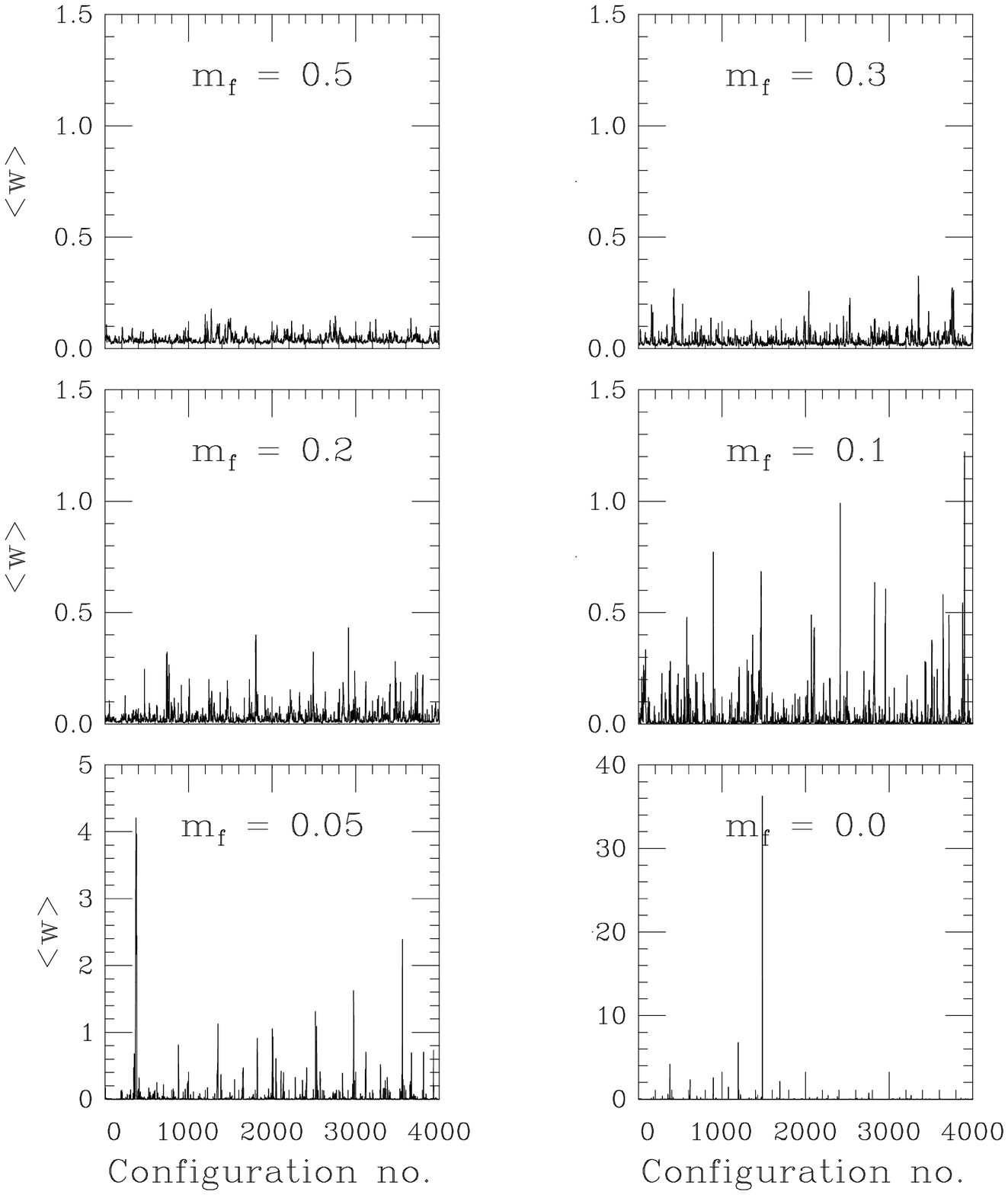}{\ten}{\figsizeC}

\vskip -0.7 truein

\figure{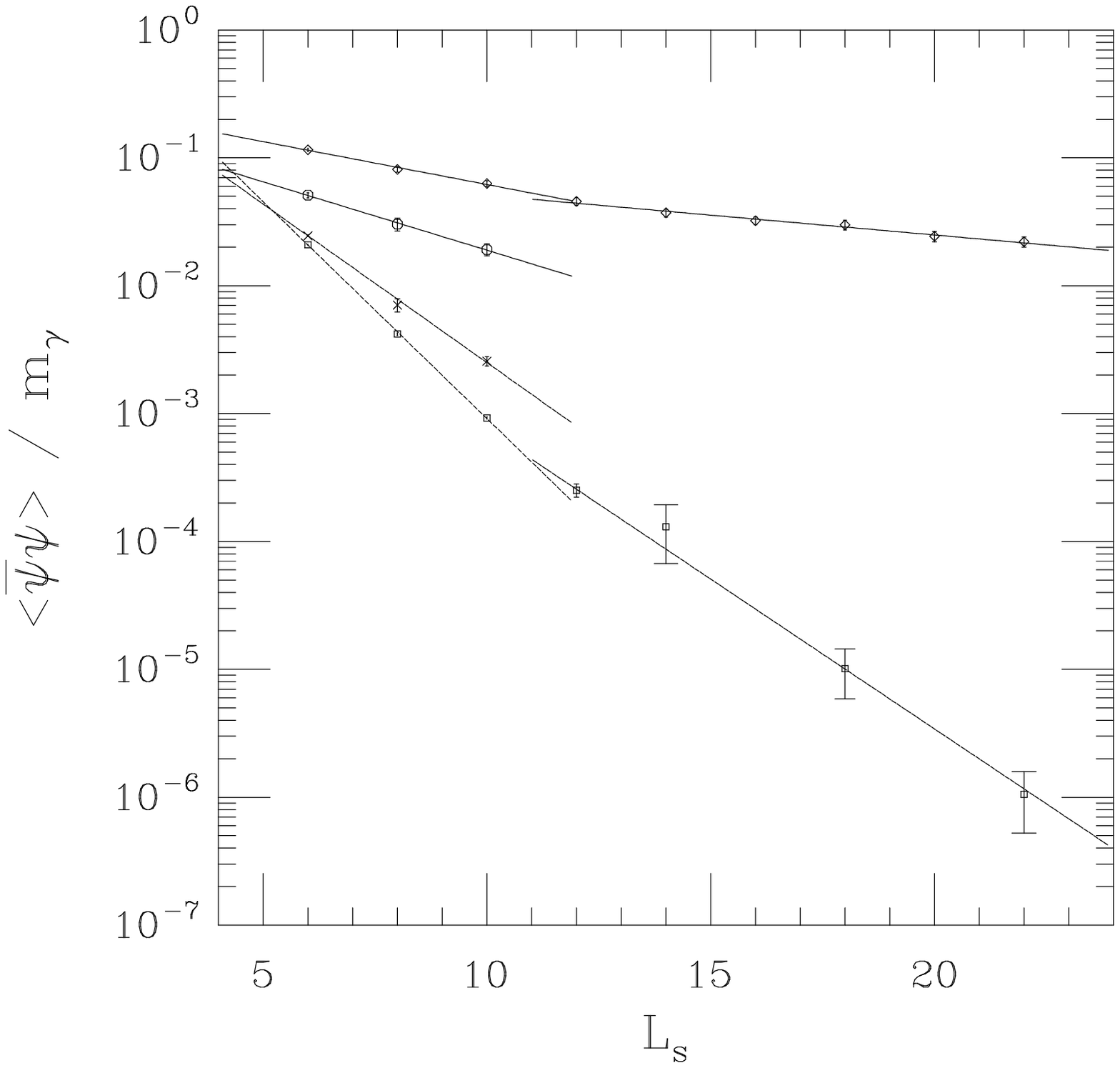}{\eleven}{\figsizeB}

\vskip -0.7 truein

\figure{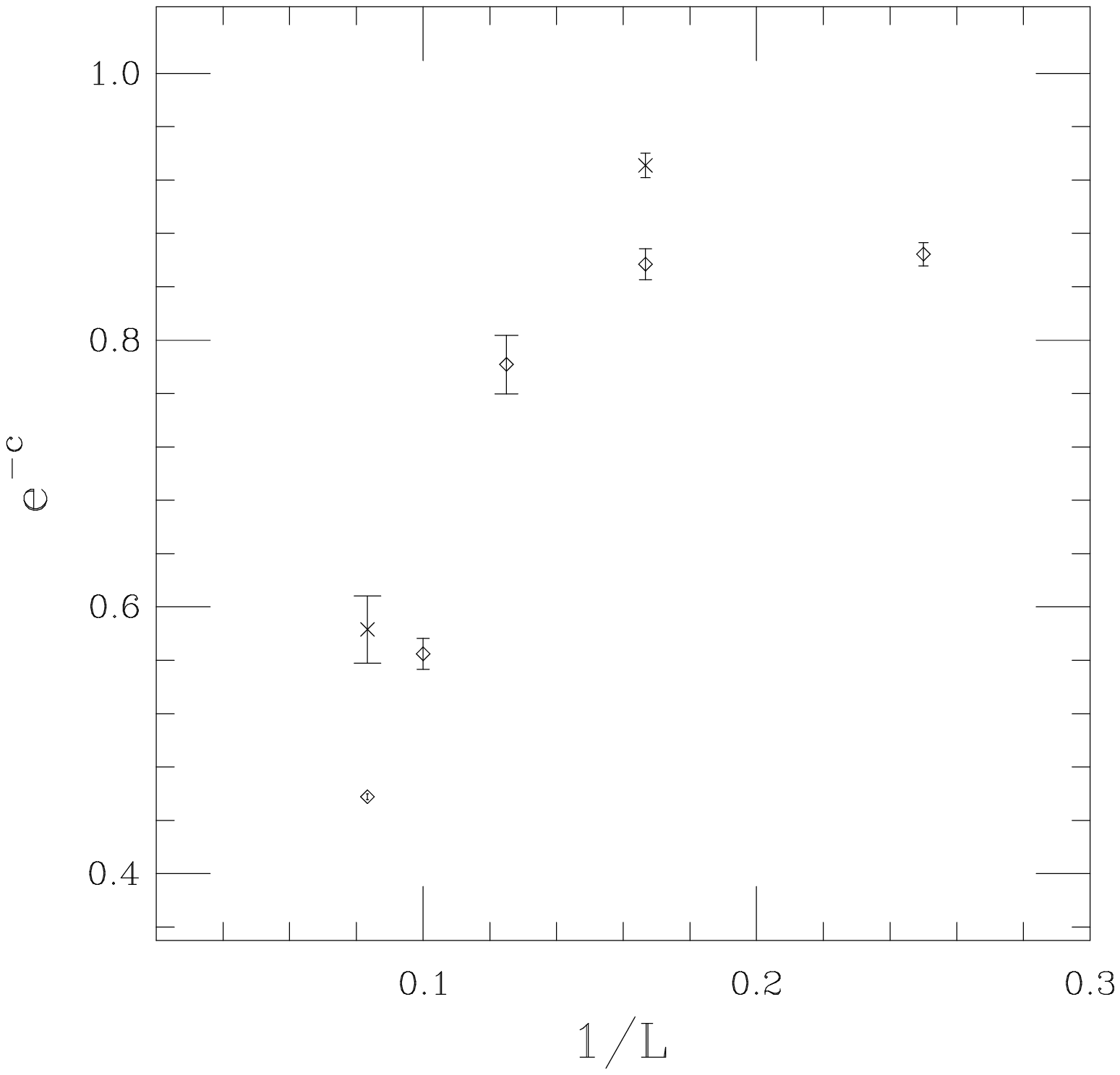}{\twoelve}{\figsizeB}

\figure{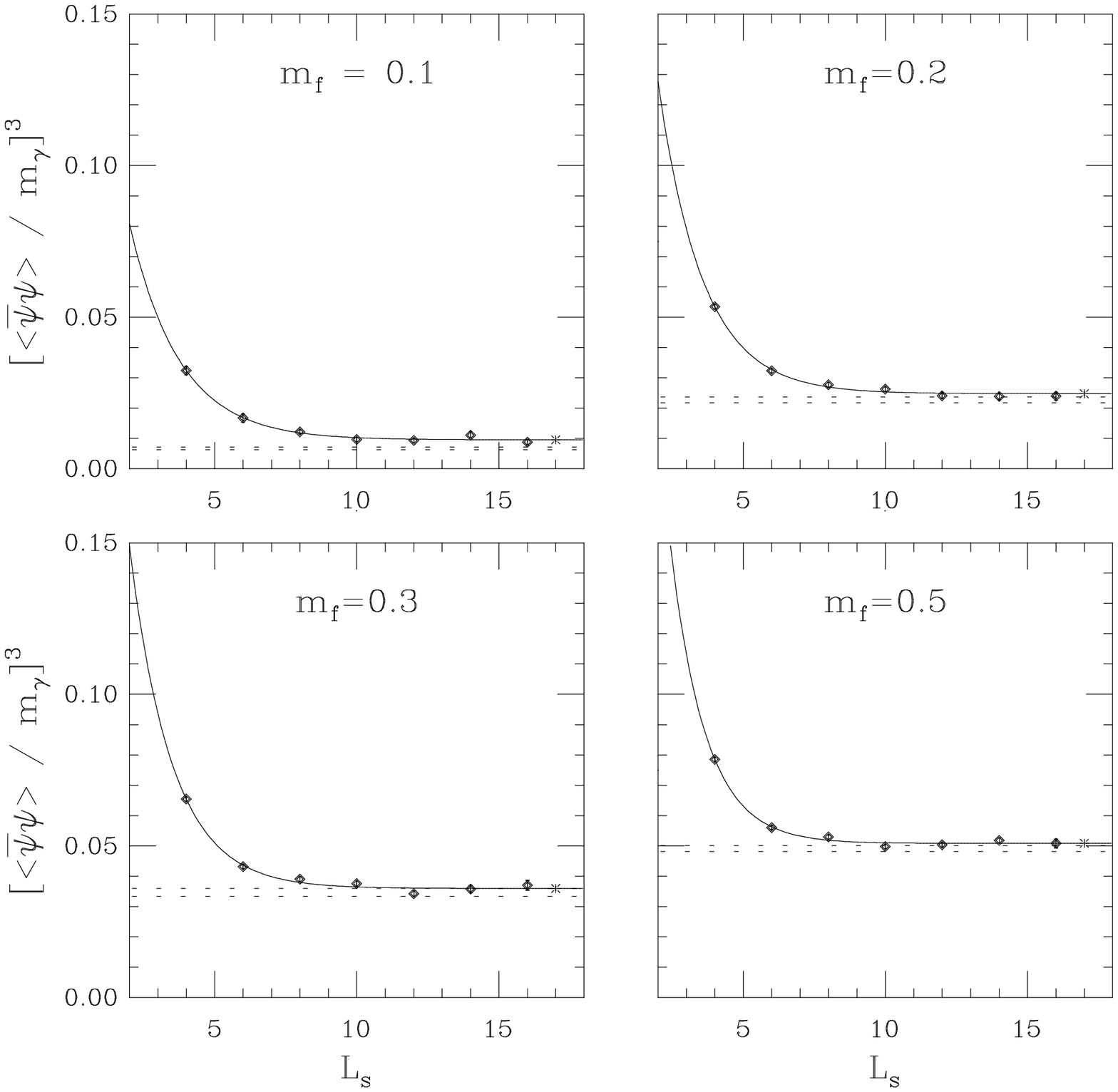}{\thirteen}{\figsizeA}

\figure{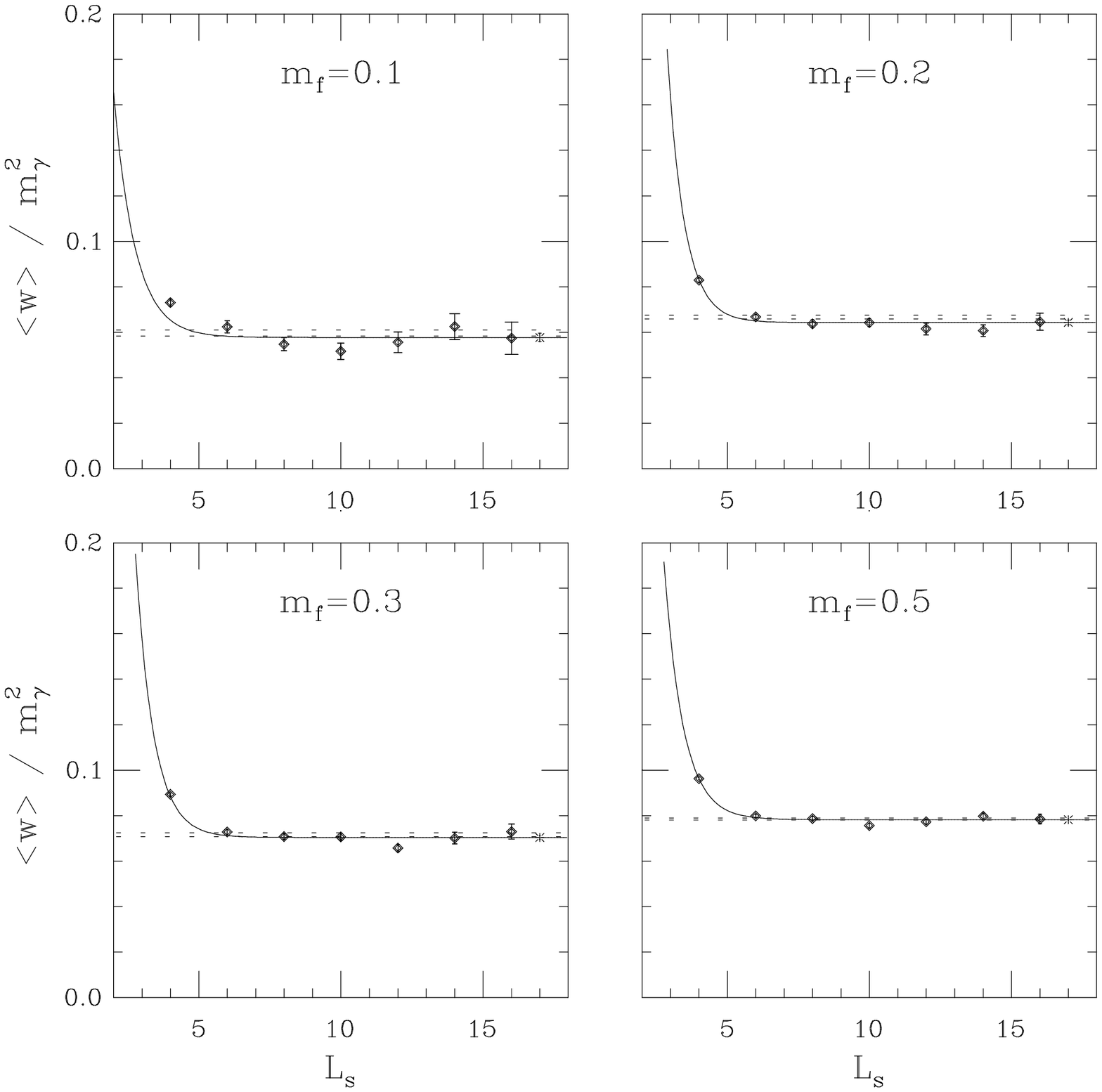}{\fourteen}{\figsizeA}

\figure{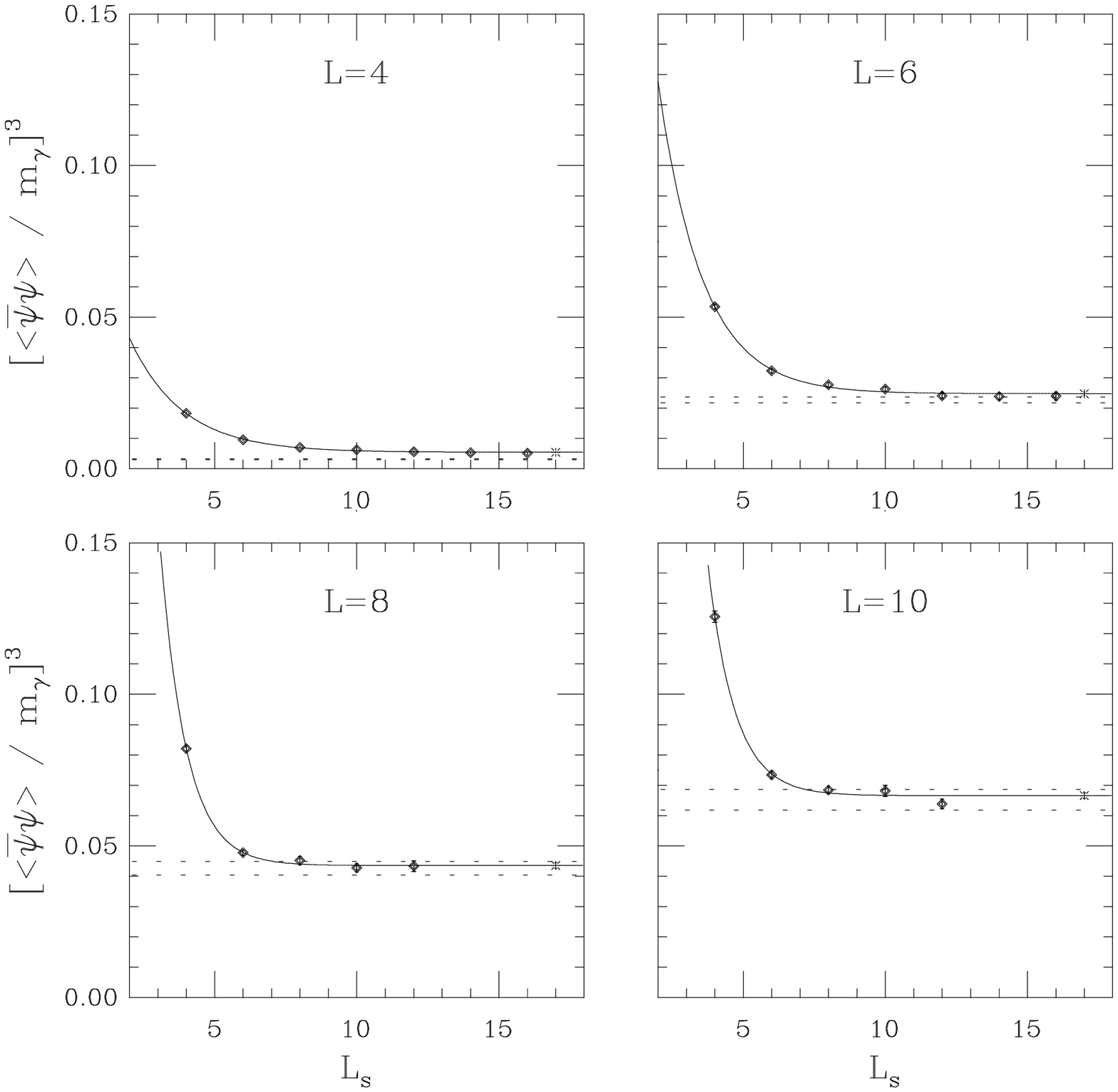}{\fifteen}{\figsizeA}

\figure{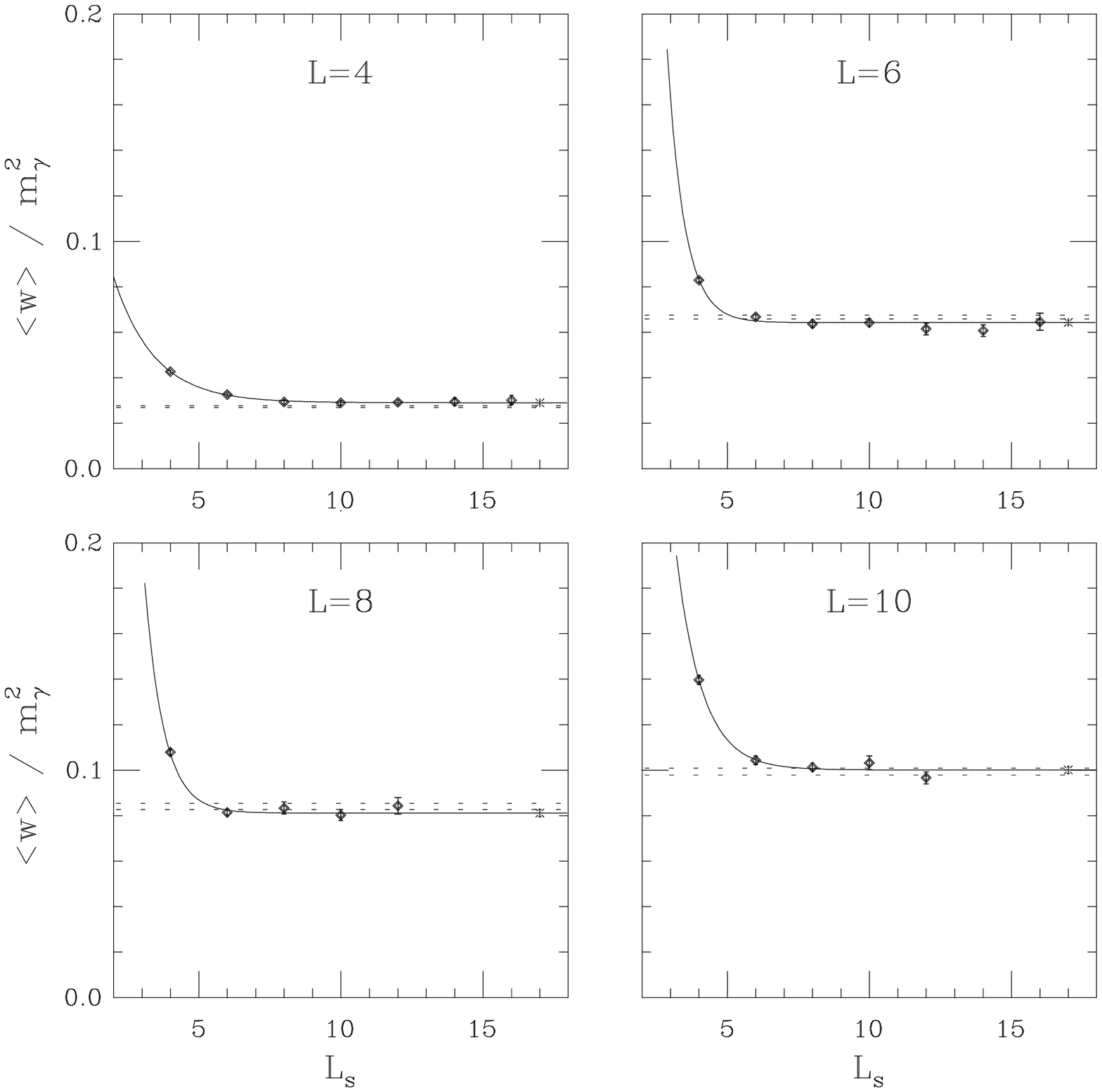}{\sixteen}{\figsizeA}

\figure{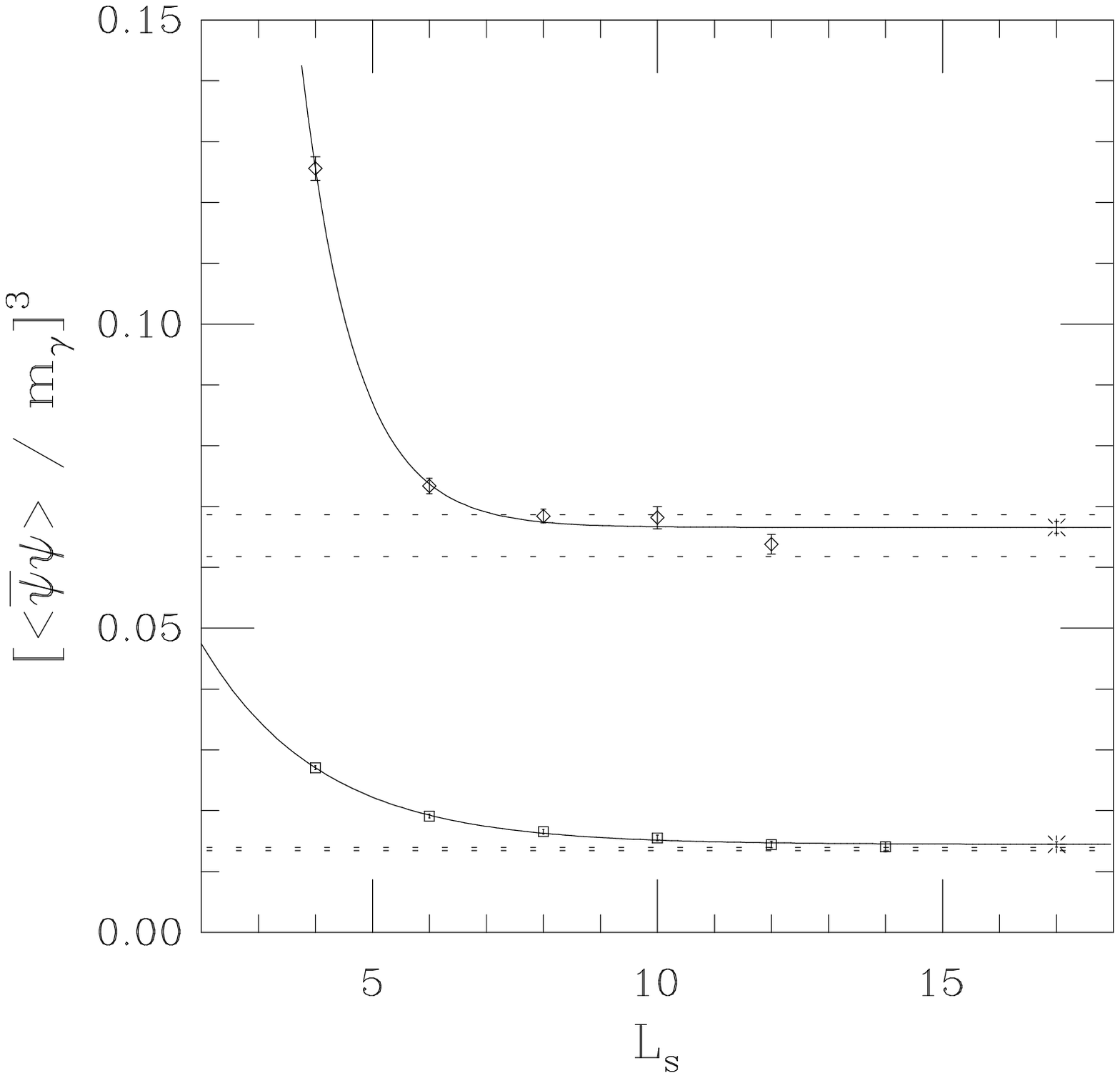}{\seventeen}{\figsizeB}

\vskip -0.7 truein

\figure{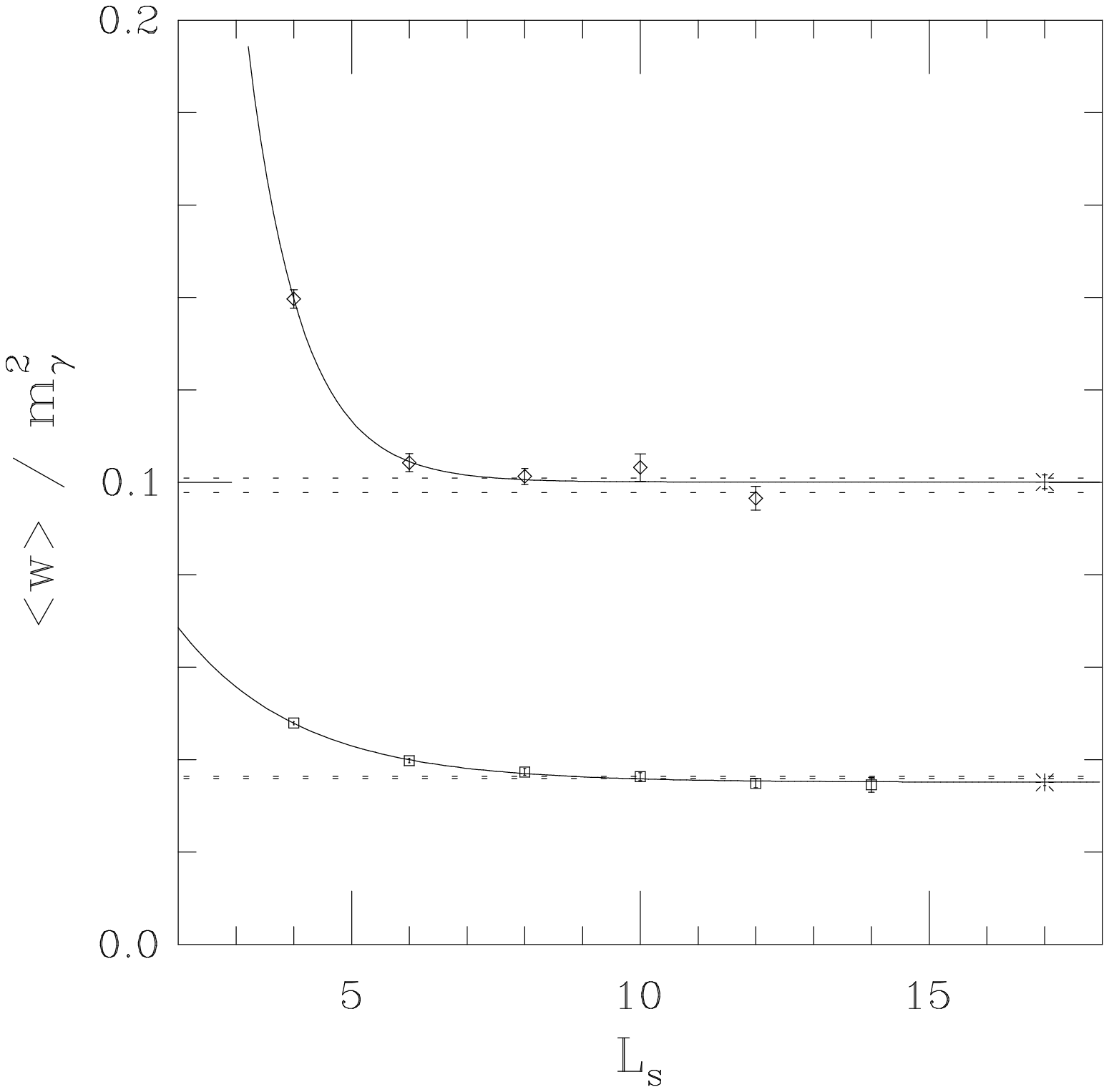}{\eighteen}{\figsizeB}

\fi


\if \epsfpreprint N 

\eject

\section* {Figure Captions.}

\noindent{\bf Figure 1:} \one

\noindent{\bf Figure 2:} \two

\noindent{\bf Figure 3:} \three

\noindent{\bf Figure 4:} \four

\noindent{\bf Figure 5:} \five

\noindent{\bf Figure 6:} \six

\noindent{\bf Figure 7:} \seven
 
\noindent{\bf Figure 8:} \eight

\noindent{\bf Figure 9:} \nine

\noindent{\bf Figure 10:} \ten

\noindent{\bf Figure 11:} \eleven

\noindent{\bf Figure 12:} \twoelve

\noindent{\bf Figure 13:} \thirteen

\noindent{\bf Figure 14:} \fourteen

\noindent{\bf Figure 15:} \fifteen

\noindent{\bf Figure 16:} \sixteen

\noindent{\bf Figure 17:} \seventeen

\noindent{\bf Figure 18:} \eighteen

\fi

\end{document}